\documentstyle[preprint,aps,floats,epsfig]{revtex}

\def\book#1[[#2]]{{\it#1\/} (#2).}
\def\am#1 #2 #3.{{\it Ann.\ Math.\ \bf#1} #2 (#3).}
\def\ap#1 #2 #3.{{\it Ann.\ Phys.\ \bf#1} #2 (#3).}
\def\apj#1 #2 #3.{{\it Astrophys.\ J.\ \bf#1} #2 (#3).}
\def\atmp#1 #2 #3.{{\it Adv.\ Theor.\ Math.\ Phys.\ \bf#1} #2 (#3).}
\def\cmp#1 #2 #3.{{\it Commun.\ Math.\ Phys.\ \bf#1} #2 (#3).}
\def\comnpp#1 #2 #3.{{\it Comm.\ Nucl.\ Part.\ Phys.\  \bf#1} #2 (#3).}
\def\cqg#1 #2 #3.{{\it Class.\ Quant.\ Grav.\ \bf#1} #2 (#3).}
\def\grg#1 #2 #3.{{\it Gen.\ Rel.\ Grav.\ \bf#1} #2 (#3).}
\def\jmp#1 #2 #3.{{\it J.\ Math.\ Phys.\ \bf#1} #2 (#3).}
\def\ijmpd#1 #2 #3.{{\it Int.\ J.\ Mod.\ Phys.\ \bf D#1} #2 (#3).}
\def\mpla#1 #2 #3.{{\it Mod.\ Phys.\ Lett.\ \rm A\bf#1} #2 (#3).}
\def\ncim#1 #2 #3.{{\it Nuovo Cim.\ \bf#1\/} #2 (#3).}
\def\npb#1 #2 #3.{{\it Nucl.\ Phys.\ \rm B\bf#1} #2 (#3).}
\def\phrep#1 #2 #3.{{\it Phys.\ Rep.\ \bf#1\/} #2 (#3).}
\def\pla#1 #2 #3.{{\it Phys.\ Lett.\ \bf#1\/}A #2 (#3).}
\def\plb#1 #2 #3.{{\it Phys.\ Lett.\ \bf#1\/}B #2 (#3).}
\def\pr#1 #2 #3.{{\it Phys.\ Rev.\ \bf#1} #2 (#3).}
\def\prd#1 #2 #3.{{\it Phys.\ Rev.\ \rm D\bf#1} #2 (#3).}
\def\prl#1 #2 #3.{{\it Phys.\ Rev.\ Lett.\ \bf#1} #2 (#3).}
\def\prs#1 #2 #3.{{\it Proc.\ Roy.\ Soc.\ Lond.\ A.\ \bf#1} #2 (#3).}

\newcommand{\be}{\begin{equation}}
\newcommand{\ee}{\end{equation}}
\newcommand{\bea}{\begin{eqnarray}}
\newcommand{\eea}{\end{eqnarray}}
\newcommand{\bml}{\begin{mathletters}}
\newcommand{\eml}{\end{mathletters}}

\begin{document}
\preprint{UFIFT-HEP-01-10}
\draft
\tighten

%\wideabs{                % Uncomment this line for two-column output
\title{Dilaton spacetimes with a Liouville potential}
\author{Christos Charmousis\footnote{E-mail address:charmousis@phys.ufl.edu}}
\address{{\footnote{Actual address: LPT, Universit\'e de Paris Sud,
B\^at. 210, 91405 Orsay CEDEX, France}}Institute for Fundamental Theory\\ 
Department of Physics,  University of Florida\\ 
Gainesville FL 32611-8440, USA}
\date{\today}
\setlength{\footnotesep}{0.5\footnotesep}
\maketitle

\begin{abstract}
We find and study solutions to the Einstein equations in $D$ dimensions 
coupled to a scalar
field source with a Liouville potential under the assumption 
of $D-2$ planar symmetry. The general static or 
time-dependent solutions
are found yielding three classes of $SO(D-2)$ symmetric 
spacetimes. In $D=4$ homogeneous and isotropic 
subsets of these solutions yield planar scalar field 
cosmologies.  
In $D=5$ they represent the  general static or 
time-dependent backgrounds 
for a dilatonic wall-type brane Universe of planar cosmological symmetry.
Here we apply these solutions as $SO(8)$
symmetric backgrounds to
non-supersymmetric 10 dimensional 
string theories, the open $USp(32)$ type I string and the heterotic
string $SO(16)\times SO(16)$. We obtain the general $SO(9)$ 
solutions as a particular case. All static solutions are found to be 
singular with the singularity sometimes hidden by a horizon. The solutions are 
not asymptotically flat or of constant curvature. The singular behavior  
is no longer true once we permit space and time dependence 
of the spacetime 
metric much like thick domain wall or global vortex spacetimes. 
We analyze  the general time and space dependent 
solutions giving implicitly a class of time and space dependent solutions 
and describe the breakdown of an extension to Birkhoff's theorem
in the presence of scalar matter. We argue that the solutions described
constitute the general solution to the field configuration under $D-2$
planar symmetry.
\end{abstract}

\section{Introduction}
The dimensional nature of the manifold we live in has been an intriguing 
question for mathematicians and physicists alike in the past century. 
Minkowski made the first crucial step \cite{mink} in this direction 
introducing time 
as part of a 4 dimensional 
space-time manifold. Of course time, at every day low velocities,
is not perceived as a coordinate, however, at accelerator velocities 
close to the speed of light, time and length can vary, the 
prefactor $c^2$ acting 
as a dimensional 
`warp-like'  factor in Einstein's theory of special relativity.
Furthermore, with the introduction of timelike coordinates 
in general relativity, dynamics are encoded in spacetime geometry. 
This geometrical idea, embodied in Mach's principle, was taken 
further by Kaluza and Klein who obtained electromagnetism as part 
of the geometry
of a {\it vacuum} 5-dimensional spacetime rather than a source term 
of the 4 dimensional Einstein equations (see for example \cite{wesson}
for a general discussion on Kaluza-Klein gravity). 

In  recent years string theory 
has been the main 
advocate of higher dimensional theories but also independently  
in the early 80's brane Universe  models 
were introduced \cite{rubi}, 
and also simple models where the 5th dimension 
was {\it dynamically} compactified by a time-like contraction 
of the 5th dimension in a 5 dimensional Kasner vacuum solution \cite{det}.
The subject was revived with H\"{o}rawa-Witten \cite{hw} 
cosmology, related works 
\cite{louk}, \cite{chamblin}, \cite{cedric}
 and in the last two years 
with the brane-Universe models 
proposed in \cite{add}, \cite{rs}.

It is reasonable to argue from the point of view of Unified  theories
that any additional dimensions of spacetime
must in some way be 
connected with string theory where gravity is unified with the 
fundamental elementary interactions. One would like to ultimately bridge 
our knowledge of standard cosmology and the 
Standard Model to the theoretical realm of string theory.
However on going to higher dimensions
one does not want to violate rather precise experimental data of standard model
physics 
and furthermore 
lose the unique characteristics of a 4 dimensional spacetime. 
Renormalisability of gauge interaction and non-trivial gravity, with 
the graviton acquiring two polarization degrees of freedom, agree only for
a 4 dimensional spacetime (for a discussion see \cite{ram}). 
Indeed the former constraint  is often 
partially embodied in the assumption 
that the Standard Model fields do not {\it see} 
the extra dimensions, being strictly confined on a 4 dimensional 'braneworld'.
Of course any fields weakly interacting with the Standard Model have no 
reason to obey such a restriction. Gravity or closed strings
in particular see the extra dimensions by definition.  
Fermionic matter such as 
sterile neutrinos can also propagate in the extra dimensions 
and models \cite{ram2} have been introduced to explain experimental data
of neutrino mass oscillations (see \cite{ram3} for a recent review on neutrino
physics). Note however that a unique sterile neutrino is less favored by the 
recent experimental data \cite{sno}. 
Furthermore cold dark matter particles such as axions or WIMPS, 
which in some cases can be 
modeled by a scalar field and a self-interaction potential, 
could also constitute such bulk matter. These rather general and 
speculative observations
are dictated to 
us by TeV scale physics.

At the theoretical realm
of beyond the standard model physics, bulk scalar matter 
originates naturally from string theory in the guise of the dilaton 
field.
It is such scalar matter, 
as a source to the Einstein equations 
that we shall be considering here. 
In supersymmetric string theories the dilaton potential is `protected'
by supersymmetry. Hence a dilaton field as matter source 
corresponds to the rather unphysical
case of a stiff perfect fluid, where pressure is equal to energy density
and the velocity of sound is the velocity of light.
The breaking of spacetime supersymmetry in 10 dimensions however, 
results in the appearance
of a dilaton tadpole, which boils down quite generically to 
 a Liouville type potential in the field 
content of the classical low energy theory.
Furthermore as a result spacetime does not admit 
solutions of maximal symmetry in 
particular a Minkowski, de-Sitter or AdS background. 
Solutions of lesser, $SO(9)$ symmetry,
 have to be 
found \cite{dm} and rather interestingly spacetime can have 
a maximally symmetric
 9-dimensional behavior, the 10th dimension being compact. 
Furthermore from a cosmology induced perspective 
the energy-momentum tensor in the presence 
of the potential and a homogeneous time-dependent scalar field
can be treated as a perfect fluid source with energy-density 
$\rho$ and
pressure $P$ given by,
$$
\rho=-{1\over 4}\partial_\mu\phi\partial_\nu\phi g^{\mu\nu}
+V(\phi)\qquad P=-{1\over 4}\partial_\mu\phi\partial_\nu\phi g^{\mu\nu}
-V(\phi)
$$
Unlike the free scalar field case no equation of state is now a priori 
specified for $V\neq 0$ (see however in this context \cite{quint1}).

In this paper we shall find 
solutions to the Einstein 
equations coupled with a scalar field and a self-interaction Liouville
 potential, namely 
we shall consider a classical action of the form,
\be
\label{action}
S_E = {1 \over 2k^2} \int d^{D} x \sqrt{-G} [ R - {1 \over 2}
(\partial \Phi )^2 -2 \alpha e^{\gamma \Phi}], 
\ee
We shall 
carry out our analysis in an arbitrary number of dimensions $D$. We shall
consider spacetimes admitting $D-2$ planar spacelike surfaces. This is
the  cosmological setting for a planar dilatonic domain 
wall brane Universe 
in $D=5$ first discussed in \cite{chamblin} (see also \cite{dil}, 
\cite{bir}, \cite{davidmaria}).
Here we shall be applying our $SO(8)$ solutions in $D=10$ tachyon free, 
10 dimensional non-supersymmetric string theories. The possibility of 
having anomaly free non supersymmetric string theories was noted in 
\cite{charles}. Tachyon free non-supersymmetric models were constructed 
in \cite{harvey} and more recently in \cite{sagnotti} and  \cite{sugimoto}
(for a recent discussion on anomaly related issues see \cite{sw}).
We shall consider solutions to the open type I string with 
gauge group $USp(32)$ \cite{sugimoto}
and the $SO(16)\times SO(16)$ closed heterotic string \cite{harvey}
with cosmological constant.
The dilaton tadpole is portrayed by the Liouville potential in (\ref{action}) 
which yields $\gamma=3/2$ for the open string and  
$\gamma=5/2$ for the heterotic string to leading order 
in the string coupling expansion. 

On considering solutions with $SO(9)$ symmetry
it is found that the presence of the dilaton induces naked singularities
of spacetime where the low energy classical theory breaks down \cite{dm}. 
One can question  the persistence of this singularity if the
symmetry of spacetime is relaxed to 8-dimensional Poincar\'e symmetry.
Indeed topological defect solutions  such as 
domain walls \cite{vil}, \cite{sikivie} are not singular (away from the 
distributional source of course) once spacetime is allowed to be time 
dependent. In complete analogy  planar thick domain walls where matter
is described by a real stationary 
scalar field in a typical double well potential 
are everywhere smooth and space-time is non-static 
(see \cite{filipe} and references within).
Global vortices  \cite{ruthgl} are in analogy 
non-singular 
 once we allow the spacetime metric to be time dependent. We will see 
a similar property arising here (see also in this context the recent 
work of Lidsey \cite{lidsey}).

Also the question of asymptotic flatness is an important issue. As we noted
no maximally symmetric solutions exist for these models. However one would 
expect that the dilaton would roll down the ``potential well'', assuming its 
vacuum value for $\phi$ going to minus infinity. It is precisely the fact
that the dilaton potential 
acquires its minimum value at a non-finite value for the scalar (minus
infinity) which
yields all solutions not asymptotically flat or  of constant curvature.
This is a characteristic of exponential potentials investigated  in
\cite{wiltshire} for the case of spherically symmetric black holes 
(for a more general discussion on properties of 
massive dilatonic black holes see 
\cite{harveyruth}).

Furthermore in a more mathematical frame of mind 
another question arises : Does a generalization of 
Birkhoff's theorem hold in 
the presence of a scalar field source with a Liouville potential?
Hence does spacetime admit an extra timelike or spacelike Killing vector 
reducing all possible $SO(D-2)$ solutions to being locally static {\it or} 
time-dependent? 
We will see here that this is not the case where a simple concrete 
example was provided recently in \cite{davidmaria} (see also \cite{bir}). 
On analyzing the two 
dimensional solutions we will see exactly how Birkhoff's theorem breaks 
down in the case of a scalar field.

In what follows we shall find the general static or time dependent solutions
i.e. solutions where our fields depend on a unique 
spacelike or timelike independent 
variable. We call these solutions {\it one dimensional} 
solutions. We shall also give implicitly some {\it two-dimensional} 
solutions analyzing a simple example of these \cite{davidmaria} and analyze
in detail the field equations in the presence or not of a Liouville potential. 

Three classes of one dimensional solutions emerge, 
distinguished by the discriminant of a second degree polynomial $f(p)$ 
and depending
essentially on two integration numerical constants $c$ and $d$. 
The constant $c$ will take three values $\pm1,0$ and will
characterize the topology of the solutions. The constant $d$
will be associated to the Weyl curvature of spacetime and in the presence of a
black hole horizon will represent the quasilocal mass (see for example 
\cite{mann} and references within).  In turn 
vanishing $d$  and hence Weyl tensor, will yield the general 
$SO(9)$ symmetric solutions which will be the maximal symmetry 
solutions for our field set-up.

The roots of the polynomial $f(p)$ will always yield  coordinate or 
naked singularities of spacetime. 
Class I 
solutions for example, the two distinct root case, 
will admit timelike and spacelike solutions, 
 as for example for $d=0$ the general $SO(9)$ solutions for 
the heterotic 
string: we will find that there exist three spacelike solutions
including the 
static \cite{dm} solution and one 
timelike solution, the  $D=10$ dimensional 
version of scalar field cosmology solutions 
in $D=4$ \cite{halliwell}. Such solutions have been 
recently revived in the context of quintessence
(see for example \cite{quint1}, \cite{quint}) and also multidimensional 
cosmological models (see \cite{zhuk} and references within).  
For Class II exactly 
half of the solutions will be compact in the $p$ direction and regular 
for finite $p$. The other half will be singular at the origin and non-compact.
Class III solutions will be always compact.
In particular we will find that the  Type I
open string holds a particular singular position
and will have to be treated separately yielding completely different results.

A general characteristic of all static solutions is that they are 
of finite proper distance i.e. of $M^9 \times S^1/Z_2$ topology if and only if 
the curvature tensor is singular at the endpoints of the interval. 

We start in the next section by setting up the field equations
we shall study. We then in section III extend the method of \cite{prc} and 
discuss black hole type solutions.
In section IV and V  we give the general static or time dependent solutions
for an $SO(D-2)$ symmetric spacetime. We analyze in detail 
the full equations in
section VI, giving implicit 2-dimensional solutions 
and also give a sketch of the general two dimensional solution in the 
absence of the potential ($\alpha=0$). 
We summarize our results and conclude in section VII.

\section{General set-up}

In this section we shall give the formal setup of the
field equations for an arbitrary number of spacetime dimensions $D$.
Consider the classical theory described by the following  
effective action (written in the Einstein frame),
\be
\label{actionD}
S_E = {1 \over 2k^2} \int d^{D} x \sqrt{-g} [ R - {1 \over 2}
(\partial \Phi )^2 -2 \alpha e^{\gamma \Phi}], 
\ee
 where $\alpha$ and $\gamma$ are positive constants of our theory.
For string theory in $D=10$ they are 
related to the string tension and the leading coefficient 
in the string coupling expansion respectively.
For example for the $USp(32)$ Type I string $\alpha=64 k^2 T_9$ where $T_9$
is the positive tension of the space filling $\bar{D}_9$ brane and 
$\gamma=3/2$ is derived by the disc and projective plane amplitudes 
of the open string theory\footnote{Note that in principle we could be 
including the next order contribution ($\gamma=5/2$) 
originating from the torus
in the string 
coupling constant expansion. We shall comment on such a possibility
in our conclusions}.

Let us consider a space-time metric admitting a $(D-2)$-dimensional
planar spacelike surface. 
A general metric admitting this 
symmetry can be written \cite{tony},
\be
\label{metricD}
ds^2=e^{2\nu} B^{-{D-3\over D-2}}(-dt^2+dz^2)+B^{2\over D-2}dx_{D-2}^2
\ee
where  $\nu$ and $B$ are functions of a timelike coordinate $t$ and 
spacelike coordinate $z$. 
Note that metric (\ref{metricD}) is the typical bulk setup for brane Universe 
cosmology \cite{prc}. 
Our source  is a scalar field 
$\phi=\phi(t,z)$ with a Liouville self-interaction potential 
given by the matter Lagrangian and
energy-momentum
tensor
read off from (\ref{actionD})
\bml
\label{sourceD}
\bea
\label{lagD}
{\cal L_M}&=&-{1 \over 2}\partial_\lambda\phi \partial^\lambda\phi-2\alpha e^{
\gamma\phi}\\
T_{ab}&=&{1\over 2}\partial_a\phi\partial_b \phi+{1\over 2}g_{ab}{\cal L_M}
\eea
\eml
We shall seek solutions to the coupled Einstein and 
scalar field equations for metric (\ref{metricD}) and 
dilaton field matter (\ref{sourceD}) in $D$ dimensions,
\bea
\label{coupledD}
R_a^b&=&T_a^b-{\delta_a^b\over D-2}  T\\
\Box\phi-2\alpha \gamma e^{\gamma \phi}&=&0.
\eea 
After some reshuffling the field equations take the form,
\bml
\label{einsteinD}
\bea
B_{tt}-B_{zz} &=& 2\alpha B^{1\over D-2}e^{2\nu+\gamma\phi}\label{wave1D}\\
\nu_{tt}-\nu_{zz} &=& \frac{\alpha}{D-2} B^{-{D-3\over D-2}} 
e^{2\nu+\gamma\phi}
+{1 \over 4}(\phi_z^2 - \phi_t^2)\label{wave2D}\\
\phi_{tt}-\phi_{zz}&=&-2\alpha \gamma B^{-{D-3\over D-2}}e^{\gamma\phi+2\nu}+
\phi_z {B_z \over B}-\phi_t {B_t \over B}\label{wave3D}\\
2\nu_{z}B_{t}+2\nu_{t}B_{z} - 2 B_{{tz}}&=& B\phi_t\phi_z
\label{int1D}\\
2\nu_{z}B_{z}+2\nu_{t}B_{t} - B_{{tt}}-B_{{zz}}&=& 
{B\over 2}(\phi_t^2+\phi_z^2)
\label{int2D}
\eea
\eml  
The metric components (\ref{metricD}) are expressed in such a way so that 
equations (\ref{wave1D}), (\ref{wave2D}) and  (\ref{wave3D}) 
are non-homogeneous wave equations with respect to $B$,  $\nu$ and $\phi$
for all $D$
whereas (\ref{int1D}) and (\ref{int2D}) are
interpreted as integrability equations for these.  If 
$\gamma=0$ and the scalar field is constant throughout 
spacetime, then the field equations 
reduce to the Einstein equations in the presence of a cosmological constant 
$\alpha$. The general solution for this case 
has been found and treated in detail 
in \cite{prc}. 
In the same frame of mind it is useful to 
transform to light 
cone coordinates,
\be 
\label{lightD}
u=\frac{t-z}{2},\qquad v=\frac{t+z}{2}
\ee 
upon which the field equations reduce to,
\bml
\label{einstein2D}
\bea
B_{uv} &=& 2\alpha B^{1\over D-2} e^{2\nu+\gamma\phi}\label{wave11D}\\
\nu_{uv} &=& \frac{\alpha}{D-2} B^{-{(D-3)\over (D-2)}} e^{2\nu+\gamma\phi}
-{1\over 4}\phi_u\phi_v\label{wave12D}\\
\phi_{uv} &=& -2\alpha \gamma B^{-{D-3\over D-2}}e^{\gamma\phi+2\nu}
-{1\over 2B}(\phi_uB_v+\phi_vB_u\label{wave13D})\\
2\nu_{u}-\left[ln(B_{u})\right]_{u} &=& {B \over 2B_u}\phi_u^2\label{int11D}\\
2\nu_{v}-\left[ln(B_{v})\right]_{v} &=& {B \over 2B_v}\phi_v^2\label{int12D}
\eea
\eml
a system of 3 non-homogeneous wave equations for 
$B$, $\nu$ and $\phi$ constrained by the {\it ordinary} 
differential equations (\ref{int11D}), (\ref{int12D}). It can be shown that 
(\ref{wave12D}) is redundant resulting  from the remaining equations, 
however, its form
can be instructive and we keep it. 

Before proceeding into the search for solutions to (\ref{einstein2D}), let 
us note the two dimensional (in the $t-z$ plane) conformal symmetries,
\be
\label{confD}
u\rightarrow f(u)\qquad v\rightarrow g(v)
\ee 
where $f$ and $g$ are arbitrary functions which 
leave (\ref{metricD}) invariant. 
This is an essential symmetry of the problem in the metric 
Anzatz we have chosen, which will be seen to reduce 
seemingly two-dimensional 
solutions, to solutions which are in fact 
one-dimensional. Also the form of the metric 
(\ref{metricD}) dictates that with little effort from a static one-dimensional 
solution we can obtain a time dependent one dimensional solution. 
We shall call one-dimensional solutions the ones that after a suitable
coordinate transformation can be seen to depend (locally)
only on a
timelike  {\it or} a spacelike coordinate.   
Two dimensional solutions will be those
depending on a timelike  {\it and} spacelike coordinate 
such that there exists no coordinate 
transformation or equivelantly no timelike or spacelike 
Killing vector reducing them to a one-dimensional solution.

\section{Dilaton black holes}

As a first approach
 let us construct solutions using the method of \cite{prc} which extend the 
topological black hole solutions \cite{kottler}. The topological 
black hole solutions have been extensively studied in string 
theory (see \cite{roberto} and references 
within) and also in brane Universe cosmology \cite{BDEL} 
the earliest application dating 
\cite{kraus} in the
context of the Randall-Sundrum \cite{rs} model. We will see in the next section
how the dilatonic version of these 
solutions are part of a more general class of solutions.

The starting point are 
conditions (\ref{int11D}) and (\ref{int12D}) which are not integrable
equations as they stand. This is the mathematical difficulty we will 
have to face throughout our analysis and will lead to the 
breakdown of Birkhoff's unicity theorem. So
choose a simple Anzatz that makes  (\ref{int11D}) and (\ref{int12D}) 
integrable equations,
\be
\label{ansatz}
e^\phi=B^c e^{\phi_0}
\ee
with c and $\phi_0$  real constants. On doing so, constraints (\ref{int11D}) 
and (\ref{int12D}) yield,
\be
\label{sol1}
B=B(U(u)+V(v)),\qquad e^{2\nu}=U'(u)V'(v)B' B^{c^2/2}
\ee
where $U$ and $V$ are arbitrary functions of a single variable 
and $'$ will always denote the derivative with respect to the unique argument
of the function.  On inputing (\ref{ansatz}),
(\ref{sol1}) 
in the wave equations of (\ref{einstein2D}) we find that they 
are consistent for $\gamma\neq 0$ if and only if
$c=-\gamma$. In order to simplify notation let us set 
\be
\label{s}
s={\gamma^2 \over 2} - {D-1 \over D-2}.
\ee
We will see that the sign of $s$ plays a particularly important role, often 
determining 
the nature of the solutions. 
The case  $c=\gamma=0$ corresponds to the 
cosmological constant solution \cite{kottler}.  
The wave equation (\ref{wave11D}) for component $B$ reduces to,
\be
\label{b}
B'=-{2\alpha \over s}B^{-s}e^{\gamma\phi_0}-d/2
\ee
and in particular for $s=0$
\be
\label{b32}
B'=2\alpha e^{\gamma\phi_0} \ln B-d/2
\ee
with $d$ an arbitrary integration constant.   
Hence (\ref{metricD}) admits the particular solution,
\be
\label{blackD}
ds^2=B'B^{{\gamma^2\over 2}-{D-3\over D-2}}U'V'(-dt^2+dz^2)+B^{2 \over D-2}
dx_{D-2}^2
\ee
for $s\neq 0$, with $B'$ given by (\ref{b})
and similarly for $s=0$ with $B'$ given by (\ref{b32}).

Now the important point is that this solution is a 
one dimensional solution. Indeed
$U$ and $V$ reflect the conformal rescaling freedom (\ref{confD}) 
and are not physical degrees of freedom.
To see this we fix (\ref{confD}) setting,
\be
\label{kruskal}
U={1\over 2}(\bar{z}-\bar{t}),\qquad V={1\over 2}(\bar{z}+\bar{t})
\ee
and hence $B=B(\bar{z})$ i.e. the solution is {\it static} 
for $\bar{z}$ spacelike and vice-versa\footnote{Alternatively take,
$U={1\over 2}(-\bar{z}+\bar{t}),\quad V={1\over 2}(\bar{z}+\bar{t})$
upon which $B=B(\bar{t})$.}. 
Noting then that $d\bar{z}={dB\over B'}$
we get,
\be
\label{marionD}
ds^2=-B'B^{{\gamma^2\over 2}-{D-3\over D-2}}
(-d\bar{t}^2)+{B^{{\gamma^2\over 2}-{D-3\over D-2}}
\over -B'} 
dB^2+B^{2\over D-2}dx_{D-2}^2
\ee
for $s\neq 0$, with $-B'$ given by (\ref{b}) as a function\footnote{
For the $\bar{t}$-case simply replace $-B'$ by $B'$ in (\ref{marionD})} of $B$
and accordingly for  $s=0$ using (\ref{b32}).
The dilaton field is given by, 
$$
\phi=\phi_0-\gamma \ln B
$$
for all values of $s$.
The form of these solutions, which are singular at $B=0$,
depends on the sign and zeros of $B'$.
In fact as we can see from (\ref{b}) and (\ref{b32})
depending on the sign of $d$ and $s$, 
we will have black hole solutions or time 
dependent solutions with a cosmological horizon or again 
solutions with a naked singularity at $B=0$. 
These solutions have been 
found and analyzed in a different coordinate system by
Chan et al \cite{mann} for the spherical case and by Cai et al \cite{cai} 
for  planar 
and hyperbolic spatial $D-2$ geometry 
(see also recent work \cite{zlo} for a general class 
of solutions). As
 backgrounds 
for the motion of dilatonic domain wall type Universes they were 
analyzed by Chamblin and Reall \cite{chamblin} and more recently in 
\cite{davidmaria}. 
Since we are interested  
in the context of non-supersymmetric 
string theories 
we examine these solutions for $D=10$ referring the reader to the above papers
for further applications and properties of these solutions.

First of all we note that $\gamma=3/2$ i.e. (\ref{s}) yields $s=0$, 
the critical value for the 
gravitational field (but not for the dilaton field), 
and corresponds
to the Type I $USp(32)$ string. Hence whatever spacetime
solutions we find for non-supersymmetric heterotic string theories, 
$\gamma=5/2$,
 the Type I string solutions will be inherently
different. 

Let us start with the case of $\gamma=5/2$ ($s=2$), the 
non-supersymmetric heterotic string case.
Then using (\ref{b}), solutions (\ref{marionD}) simplify to,
\be
\label{het}
ds^2=\eta (-\alpha e^{5\phi_0/2}B^{-2}-d/2)B^{9/4}(-d\bar{t}^2)+
{B^{9/4}dB^2\over \eta (-\alpha e^{5\phi_0/2}B^{-2}-d/2 )}+B^{1/4}dx_8^2
\ee
where $\eta=\pm1$, with 
$$
\phi=\phi_0-{5\over 2} \ln B
$$
Solutions (\ref{het}) can be written in the string frame by the 
Weyl rescaling, $g_{\mu\nu}^{(S)}=e^{\phi/2}g_{\mu\nu}$.
The $\eta=-1$ corresponds to (\ref{marionD}) and the $\eta=1$ to its 
counterpart.
Hence for $\eta=-1$ if $d<0$ then 
$-B_H'=\alpha e^{5\phi_0/2}B_H^{-2}+d/2=0$ 
corresponds to a cosmological horizon $B_H$ and
hence for the horizon exterior,  
$B$ is a timelike coordinate with $B=0$  a timelike singularity. 
For $d>0$, $B$ is 
spacelike, and $B=0$ is a naked  timelike singularity. 
Alternatively for $\eta=1$ if $d<0$ then $B_H'=0$ corresponds to a black hole
horizon and hence for $B>B_H$, $B$ is a spacelike 
coordinate and $B=0$ is now a spacelike singularity. The parameter $d$ in 
this case is related to the quasilocal mass (see for example \cite{mann} and 
references within) of the black hole.  
For $d>0$, $B$ is timelike with $B=0$ a naked spacelike singularity.

Let us now turn to the asymptotic behavior.
Calculation of the Ricci scalar gives, 
$R={5\over 8}(-9\alpha+5-d/2 B^2)B^{-25/4}$
which ties in with the  form of the 
dilaton potential. Note then 
that spacetime geometry coupled to 
matter is well behaved 
for large $B$, the Ricci scalar asymptoting zero.
On the other hand  note now the bizarre property of these
solutions; their asymptotic behavior depends on the ``mass'' parameter 
$d$. Indeed for large 
$B$ the solution asymptotes,
$$
ds^2\sim -\eta d B^{9/4}(-d\bar{t}^2)+
{2B^{9/4}dB^2\over -\eta d }+B^{1/4}dx_8^2
$$
and spacetime  is not asymptotically flat although matter is vanishing 
at that region. 
Indeed as our coordinate $B$ goes to infinity 
the dilaton field goes to minus infinity, the dilaton potential acquiring 
thus its ``global'' minimum. Hence even though the scalar 
field rolls down the potential well spacetime
is not of trivial curvature. 
The (partial) resolution of this discrepancy  
lies in the physical interpretation
 of $d$ which is related to the Weyl tensor.
Indeed setting $d=0$ one can show that the Weyl tensor is identically zero. 
Hence we can deduce that although matter tends to the vacuum 
(and the Ricci tensor with it) the Weyl tensor i.e. pure curvature 
controls the large $B$ region where spacetime exhibits purely 
gravitational tidal forces  
depending on the magnitude of $d$. So a far away observer for the dilatonic
black hole 
is subject to  gravitational tidal 
forces dependent on the quasilocal mass $d$, 
quite in contrast to the standard situation of a Schwarzschild black
hole which is asymptotically flat regardless of its ADM mass. 

Setting $d=0$ we get solutions of 
maximal $SO(9)$ symmetry for the heterotic string, $\gamma=5/2$
which were first discussed in \cite{lupope}.
The solution reads,
\be
\label{pope}
ds^2=B^{1/4}[-\eta d\bar{t}^2+dx_8^2+ 
{\eta B^{4}dB^2\over \alpha e^{5\phi_0/2}}]
\ee
and exhibits a naked singularity at $B=0$.
Furthermore as we pointed out for 
$d=0$ spacetime is conformally flat, $ds^2\sim w^{1/12}(\eta_{ab}dx^a dx^b)$
where $w=B^{3}$. Note that even now  the solution is not asymptotically 
flat. This following the works of Poletti and Wiltshire \cite{wiltshire} for 
spherically symmetric scalar black holes is 
due to the fact that the potential attains its minimum value only at 
an infinite value for the scalar field $\phi$. The only cases where asymptotic
flatness or constant curvature is obtained is $\gamma=0$ and also by 
considering 
$\gamma=1/2$ in the string frame.

For the particular case of $\gamma=3/2$ i.e. the Type I string we obtain,
\be
\label{black32}
ds^2=B^{1/4}\left[\eta(2\alpha e^{3\phi_0/2}\ln B-d/2) (-dt^2)+{dB^2\over 
\eta(2\alpha e^{3\phi_0/2}\ln B-d/2)}
+dx^2_8\right]
\ee
$$
\phi=\phi_0-{3\over 2}\ln B
$$
In this case the sign of $d$ does not affect the solution.
We will always get a black hole horizon with  $\eta=1$  and a 
cosmological horizon with  $\eta=-1$ situated at $B'_H=0$. Spacetime 
is singular at $B=0$ as
the Ricci scalar is given by,
$$
R=\pm{1\over 8}{-20\alpha+18\alpha \ln(B)+9-d/2\over B^{9/4}}
$$
Again for large $B$ the curvature scalar $R$ vanishes signaling the good 
behavior of the spacetime geometry coupled to matter. However the 
Weyl curvature  now turns out to be dependent on $\alpha$. 
Since $\alpha\neq 0$ solutions  {\it cannot} be 
conformally flat.  
Hence again the dilaton potential rolls down to its vacuum value 
at minus infinity but spacetime is 
not asymptotically flat or of constant curvature.

\section{General one-dimensional Solutions}

The above is only one subset of the static solutions one can find. In 
this section, 
we shall extract the general one dimensional static or time-dependent solutions
to (\ref{coupledD}).
Since we are looking for 1-dimensional solutions we will either have 
${\cal{A}}_u={\cal{A}}_v$
for time dependent or ${\cal{A}}_u=-{\cal{A}}_v$ for static solutions, 
where 
${\cal{A}}$ stands for
$B$, $\nu$ and $\phi$. It suffices in view of the form of our metric 
to find the static solutions, so let us concentrate on this case.
The field equations are written,
\bml
\label{static}
\bea
B'' &=& -2\alpha B^{1/(D-2)} e^{2\nu+\gamma\phi}\label{static1}\\
\nu'' +{1\over 4}\phi'^2&=& -\frac{\alpha}{D-2} B^{-(D-3)/(D-2)} 
e^{2\nu+\gamma\phi}
\label{static2}\\
\phi'' +{B'\over B}\phi'&=& 2\alpha\gamma B^{-(D-3)/(D-2)} e^{2\nu+\gamma\phi}
\label{static3}\\
2B'\nu'-B'' &=& {B \over 2}\phi'^2\label{static4}
\eea
\eml
where $'$ denotes the derivative with respect to z. Note the sign 
ambiguity of the field equations in that the 
t-dependent equations are obtained by simply replacing $\alpha$ by $-\alpha$
in (\ref{static}). 
Now we make use of the similar form of the non-homogeneous 
terms in the wave equations 
i.e. using (\ref{static1}) we can integrate  (\ref{static3}),
\be
\label{static22}
\phi'={1\over B}(c-\gamma B')
\ee
and then in turn, (\ref{static1}), (\ref{static4}) in (\ref{static2}) yield
after direct integration,
\be
\label{static23}
2\nu'={1\over B}(d+ {D-1\over D-2}B')
\ee
with $c$ and $d$ arbitrary integration  constants.
From (\ref{static22}) we see 
that $c=0$ corresponds to the dilaton solution discussed in 
the previous section and hence $d$ is again the Weyl parameter.
Replacing now the above expressions for $\phi'$ and $\nu'$ in (\ref{static4})
we get a second order differential equation for $B$,
\be
\label{B}
(d+\gamma c){B'\over B}-s{B'^2\over B}- B''-{c^2\over 2B}=0
\ee
where $s$ is given by (\ref{s}).
This equation is consistent modulo a constant 
with the remaining equation (\ref{static1}) and this can be 
shown by differentiating both and comparing. We will make 
use of this fact later on to fix our constants and eventually determine 
if our solution is timelike or spacelike. All we  need to do 
now is solve (\ref{B}) and then find $\phi$
and $\nu$ by direct integration in some adequate
coordinate system. Equation (\ref{B}) is autonomous hence we 
coordinate transform
our metric setting $B'=p$ and hence 
$B''={dp\over dB}p$. The above equation then boils down to,
\be
\label{p}
p{dp\over dB}B=-sp^2+(d+\gamma c)p-{c^2\over 2}
\ee
and hence for $s\neq 0$
\be
\label{mouradus}
-slnB(p)=\int{pdp\over f(p)}
\ee
where $f(p)$ is a second degree polynomial of $p$ given by,
\be
\label{f(p)}
f(p)=p^2 -{d+\gamma c\over s}p+{c^2\over 2s} 
\ee
Depending on the type of 
roots of this polynomial we shall obtain  
different solutions. The roots of the polynomial will generically 
correspond to candidate singularities of spacetime. For example 
we will see that for $c=0$ one of the roots is merely a coordinate
singularity
whereas the second root is a curvature singularity, hence
 the black hole solutions of the previous section. 
 
Furthermore noting that $dz=dB/p$ 
the fields $\phi$ and $\nu$ are directly integrated as functions of $p$ 
to give,
\bml
\label{phinu}
\bea
\phi(p)&=&lnB^{-\gamma}-{c\over s} \int{ dp'\over f(p')}\\
2\nu(p)&=&lnB^{(D-1)/(D-2)}-{d\over s}\int{ dp'\over f(p')}
\eea
\eml

Since all our fields are given as functions of $p$ we coordinate transform 
using (\ref{p}) whereby $dz^2={B^2 dp^2\over s^2 f(p)^2}$ and hence,
\be
\label{generalform}
ds^2=e^{2\nu} B^{-{D-3\over D-2}}(-dt^2+{B^2 dp^2\over s^2 f(p)^2})
+B^{2\over D-2}dx_{D-2}^2
\ee

We start by analyzing the roots of the polynomial $f(p)$. 
Its discriminant is given by,
$$
\Delta={1\over s^2}[(d+\gamma c)^2-2c^2 s]
$$
Two possibilities arise according to the sign of $s$. If $s>0$
then we can have two distinct real roots $p_1$ and 
$p_2$, one double root and no real roots. For 
$s<0$ we {\it always} have $\Delta\geq 0$.
For $D=10$ we note again the particular role played by $\gamma=3/2$, 
mapping $s$ to the origin.
Indeed from (\ref{p}), if $s=0$ then we have a first degree polynomial.
 For $\gamma<3/2$ we will always 
have 2 real roots.  For $\gamma>3/2$ and hence for the heterotic string, 
$\gamma=5/2$,  
the whole spectrum of possible solutions will be permitted. 
So depending on the nature 
of the discriminant $\Delta$ we classify our solutions to Class I 
(two distinct real roots),
 Class II (double root) and Class III (imaginary roots)
solutions. We also have to examine the case $s=0$ separately.

\subsection{Class I solutions}

In this case the two distinct real roots of $f(p)$ are given by,
$p_{1,2}={d+\gamma c\over 2s}\pm {p_0\over 2}$
with $p_0=\sqrt{\Delta}>0$ and we choose $p_2<p_1$. The Class I fields 
are obtained directly from 
 (\ref{mouradus}) and then in turn from (\ref{phinu}),
$$
B(p)=B_0 {q(p)^{p_1}\over |p-p_2|^{1\over s}},
\qquad
e^{2\nu}=e^{2\nu_0} {q(p)^{d+p_1{D-1\over D-2}}\over |p-p_2|^{D-1\over
s(D-2)}},
 \qquad e^\phi=e^{\phi_0} q(p)^{c-\gamma p_1}|p-p_2|^{\gamma\over s} 
$$
where $B_0$, $\nu_0$ and $\phi_0$ are constants of integration and,
$$
q(p)=\left|{p-p_2 \over p-p_1}\right|^{1\over s p_0}
$$ 
The next step is to relate $\alpha$ to the 
integration constants: noting from (\ref{B})
that $B''B=-sf(p)$ and replacing the above solutions into
(\ref{static1}) we get
\be
\label{signature}
f(p)s=2\eta \alpha e^{2\nu_0+\gamma\phi_0}B_0^{D-1\over D-2}|f(p)|
\ee
Here $\eta=1$ means that $p$ is spacelike and $\eta=-1$ that $p$
is timelike taking care of the sign ambiguity of the field equations 
(\ref{static}).
Therefore for $s>0$ we will have spacelike solutions for  $p>p_1$ or 
$p<p_2$ whereas timelike solutions will be obtained in the finite interval 
$p_2<p<p_1$. For $s<0$ the situation is interchanged.
Once the sign, and hence the signature required, is determined the integration 
constants are related by,
\be
\label{constants}
|s|=2\alpha e^{2\nu_0+\gamma\phi_0}B_0^{D-1\over D-2}
\ee

Let us explicitly write and analyze 
the solution for $D=10$ and $\gamma=5/2$ ($s=2$). All $s>0$ solutions will 
have the same behavior. The solutions then are backgrounds to the 
non-supersymmetric heterotic string.

Now $p_1$ and $p_2$ are singular points for the fields and they either 
stand for curvature singularities or coordinate singularities.
For spacelike $p$ such that $p>p_1$ we perform a change of 
origin, $p-p_1\rightarrow p$ and use (\ref{constants}) to fix the constants. 
After some algebra the solution is written,
\be
\label{class12}
ds^2= {(p+p_0)^{{p_1-p_0\over 8p_0}}\over p^{{p_1\over 8p_0}}}\left[
-\left({p+p_0\over p}\right)^{d\over 2p_0}dt^2+dx_8^2+
 \left({p+p_0\over p}\right)^{{d\over 2p_0}+{p_1\over p_0}} {dp^2\over
4\alpha e^{5\phi_0/2} (p+p_0)^{3}p^{2}} 
\right]
\ee
\be
\label{dilaton12}
\phi=\phi_0+\left({c\over 2p_0}-{5p_1\over 4p_0}+{5\over 4} \right)\ln (p+p_0)
+\left({5p_1\over 4p_0}-{c\over 2p_0} \right)\ln p
\ee
where  $p>0$.\footnote{The spacelike solutions with
 $p<p_2$ are obtained from (\ref{class12}) by interchanging $p$ by $p+p_0$.}  
and therefore the $p=-p_0$ singularity is 
never attained for a positive $p$. 
As it stands now Class I solutions depend on two integration 
parameters $c$ and $d$. The case $c=0$ was treated 
in the previous section where 
upon making the coordinate transformation  
$p=\alpha B^{-2}+d/2$ we obtain the dilaton black hole solutions
 solutions (\ref{het}).
What is noteworthy here is that $p=0$ is no longer a singular point 
of spacetime,
it is now a horizon screening the singularity at the second root of the 
polynomial $f(p)$. 
Indeed the singular part 
of the dilaton in (\ref{dilaton12}) drops out for $c=0$.  Alternatively 
we can gauge away the $c$-dependence modulo the sign. The 
constant $c$ plays then the
 role of a topological index characterizing the dilaton
field and we shall obtain 
different solutions for $c=0,\pm 1$.

So let us now suppose that $c\neq 0$. Then consider $p\rightarrow p_0 p$ 
relabel $d$ by $d/c$ and set $\eta=|c|/c$,
\be
\label{cl1}
ds^2=[p(p+1)]^{-1/16}\left({p+1\over p }\right)^{{\eta (d+5/2)\over 16\chi}}
\left[
-\left({p+1\over p}\right)^{{\eta d\over \chi}}dt^2
+dx_8^2
+  { \left({p+1\over p}\right)^{{\eta (3d+5/2)\over 2\chi}} dp^2\over
4\alpha e^{5\phi_0/2} [(p+1)p]^{5/2}} 
\right]
\ee
\be
\label{dilcl1}
\phi=\phi_0+{5\over 8}\ln [p(p+1)]+
{5\eta (d+9/10)\over 8\chi}\ln \left[p\over p+1\right]
\ee
where $\eta=\pm 1$, $\chi= [(d+9/2)(d+1/2)]^{1/2}$ and from positivity of
the discriminant we have
 $d\in (-\infty,-9/2) \, \cup \,(-1/2,+\infty)$. Notice how (\ref{cl1}), 
(\ref{dilcl1}) are symmetric under $p\leftrightarrow p+1$ and 
$\eta=1\leftrightarrow \eta=-1$.
We can portray (\ref{cl1}) in the string frame via the 
conformal rescaling $g^{(S)}=e^{\phi/2}g$ 
and it turns out that they share the same features 
as the Einstein solutions (\ref{cl1}).

We now determine the nature of the singularity for 
$p=0$ in (\ref{cl1}). 
A direct calculation of the Ricci scalar gives,
\be
\label{ricciscalar1}
R\sim [p(p+1)]^{9/16}\left[{p\over p+1}\right]^{25\eta(d+9/10)\over 16\chi} 
\ee
and hence $R$ blows up at $p=0$ if and only if 
 $\eta(d+9/10)<0$ (calculation of 
$R_{ab}R^{ab}$ and $R_{abcd}R^{abcd}$ yields the same result). Note
from (\ref{dilcl1}) that the dilaton potential exhibits the same 
behavior and furthermore the Ricci scalar 
is always singular for large $p$. The behavior of the curvature tensor is 
related to the topology of the solution. 
Indeed consider the proper distance in the $p$ direction
 defined as,
$$
\int_0^\infty dp\, \sqrt{g_{pp}}
$$
It is easy to 
see from (\ref{cl1}) 
that proper distance is finite if and only if, $\eta(d+9/10)<0$ i.e. given the
interval $d\in (-\infty,-9/2) \, \cup \,(-1/2,+\infty)$ 
we have that $d<-9/2$ for $\eta=1$,
and $d>-1/2$ for $\eta=-1$ yield a compact $p$ direction.  Hence 
we deduce that 
the $p$-direction is compact and our solution has the topology of 
an interval times a 9-dimensional manifold 
if and only if the Ricci tensor blows up at $p=0$. Hence compactness 
in the $p$-direction 
yields a singular behavior of spacetime with two naked singularities at $p=0$
and at $p\sim \infty$ (the singularity at infinity is now at a finite 
proper distance). For large $p$ spacetime takes the approximate form,
$$
ds^2\sim p^{-1/8}[\eta_{\mu\nu}dx^\mu dx^\nu+p^{-5}{dp^2\over 4\alpha 
e^{5\phi_0/2}}]
$$
and the dilaton 
$$
\phi\sim \phi_0+{5\over 4}\ln p
$$
which is just the $SO(9)$ solution (\ref{pope}). 

\begin{figure}
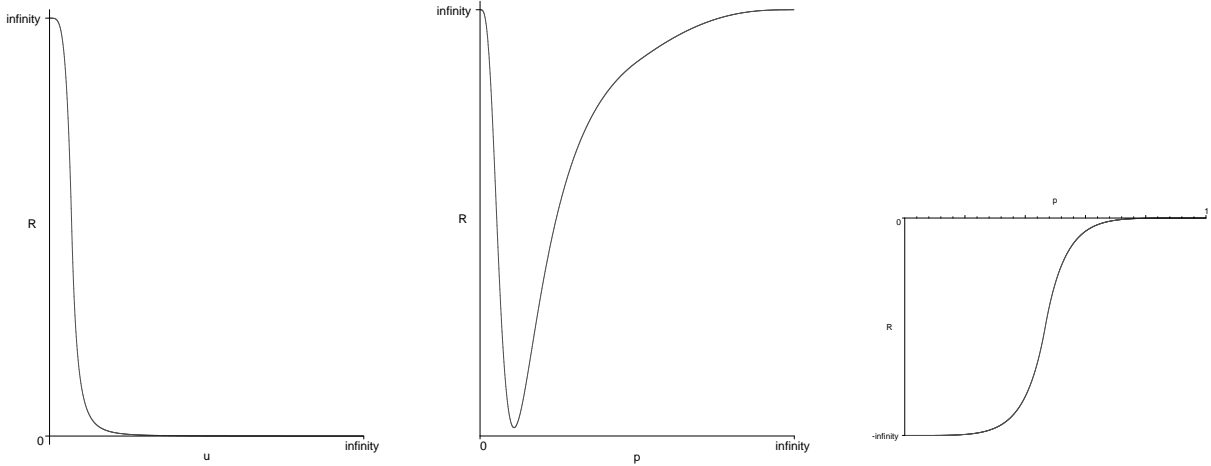

\center{\epsfig{file=cl1inf101.eps,width=5cm}\hfill 
\epsfig{file=cl1com101.eps,width=5cm}\hfill\epsfig{file=cl1time101.eps,
width=5cm}}
\vskip 5mm
\caption{Plots of the Ricci scalar for the numerical value $d=-5$. From
left to right for spacelike 
Class I   $\eta=-1,+1$ and, timelike Class I with $\eta=-1$ respectively}
\label{cl1inf1}
\end{figure}

Alternatively  if proper distance
is infinite then $p=0$ is  a coordinate singularity. It turns out 
that this solution (unlike the case $c=0$) 
cannot be extended in the timelike region so 
we make a coordinate transformation 
making $p=0$ spatial infinity, $p=1/u$. Then again we find that 
the solution is not 
asymptotically flat. From (\ref{dilcl1}) we get,
$$
\phi=\phi_0+{5\over 8}\ln{1+u\over u^2}+
{5\eta (d+9/10)\over 8\chi}\ln {1\over 1+u}
$$
and we see that the dilaton approaches minus infinity 
for large $u$ rolling down the potential well given that $\eta (d+9/10)>0$ 
(see fig \ref{cl1inf1} for a plot of the Ricci scalar{\footnote{We 
have set $\phi_0=0$, $\alpha=1$ for all the subsequent figures.}}). 

Let us now turn to the time-dependent solutions. As we mentioned above 
(\ref{signature})
in this case we have $f(p)<0$ and therefore $p_2<p<p_1$ varies in a finite 
interval. Choosing $p_2$ as our origin we note that
$dt^2={B^2dp^2\over s^2f^2(p)}$. The time dependent solution is effectively
obtained by considering solution (\ref{class12}) and replacing,
$p+p_0\rightarrow p$ and $-dt^2\rightarrow dz^2$. 
Then proceeding 
as for the spacelike case the timelike solutions for $c\neq 0$ simplify to,
\be 
\label{t1}
ds^2=[p(1-p)]^{-1/16}\left({p\over 1-p }\right)^{{\eta (d+5/2)\over 16\chi}}
\left[
-  { \left({p\over 1-p}\right)^{{\eta (3d+5/2)\over 2\chi}} dp^2\over
4\alpha e^{5\phi_0/2} [(1-p)p]^{5/2}}
+\left({p\over 1-p }\right)^{{\eta d\over \chi}}dz^2
+dx_8^2
 \right]
\ee
\be
\label{t1dil}
\phi=\phi_0+{5\over 8}\ln [p(1-p)]+
{5\eta (d+9/10)\over 8\chi}\ln \left(1-p\over p\right)
\ee
where $p$ is now a timelike coordinate and $0<p<1$ {\footnote {To pass
to an infinite timelike coordinate set 
for instance $\tilde{t}={p\over 1-p}$}}. The Ricci scalar, obtained 
by (\ref{ricciscalar1}) with $p+1\rightarrow p$ is singular at one 
of the two extremities of the interval $p=0$ or $p=1$. Indeed for $\eta=-1$
say, 
$p=1$ is a coordinate singularity and $p=0$ is a naked 
singularity. Changing the sign of $\eta$ does not produce a new solution 
but merely changes 
the direction of time. The direction of time is determined
by demanding that the scalar field tends asymptotically to minus infinity. 
We should also note  
that proper time is always infinite 
although our coordinate interval for $p$ is finite.

The form of the metric (\ref{cl1}) dictates that 
for $d=0$ i.e. for vanishing Weyl tensor  we will get the general 
solution with $SO(9)$
symmetry.

\noindent
{\bf \underline{d=0}}: Under this assumption we have,
two distinct cases depending on the sign of $\eta$. For $\eta=-1$
we will have finite proper distance in the $p$ direction and two
 naked singularities at $p=0$
and at coordinate infinity
whereas for $\eta=1$ the proper distance is infinite with $p=0$ merely a 
coordinate singularity.
Explicitly for  $\eta=-1$ we obtain from (\ref{cl1}) and 
(\ref{dilcl1}) the solution,
$$
ds^2=(p+1)^{-1/6} p^{1/24}\eta_{\mu \nu}
dx^\mu dx^\nu+(p+1)^{-7/2} 
p^{-13/8}{dp^2 \over 4\alpha e^{5\phi_0/2}}
$$
$$
\phi=\phi_0+\ln p^{1/4}+\ln (p+1)
$$
This is the $SO(16)\times SO(16)$ \cite{dm} solution as can be seen by
considering $p=ch^2(\sqrt{\alpha}y)$,
where it was noted that the solution has an effective 
 9-dimensional behavior with
the 9 dimensional Planck and Yang-Mills couplings finite. 

The  case $\eta=1$ can be obtained explicitly 
by replacing $p$ by $p+1$ and vice-versa. 
In this case spacetime is well behaved at $p=0$ the point in question 
portraying a coordinate singularity and hence as before we take $u=1/p$,
$$
ds^2=u^{1/8} (u+1)^{1/24}\eta_{\mu \nu}
dx^\mu dx^\nu+(u+1)^{-13/8} 
u^{9/8}{dp^2 \over 4\alpha e^{5\phi_0/2}}
$$
$$
\phi=\phi_0-{5\over 4} \ln u + \ln (u+1)
$$
We now have a naked singularity at $u=0$ and dilaton matter tends to 0 
for $u$ large with $\phi$ going to minus infinity. 
  
\begin{figure}
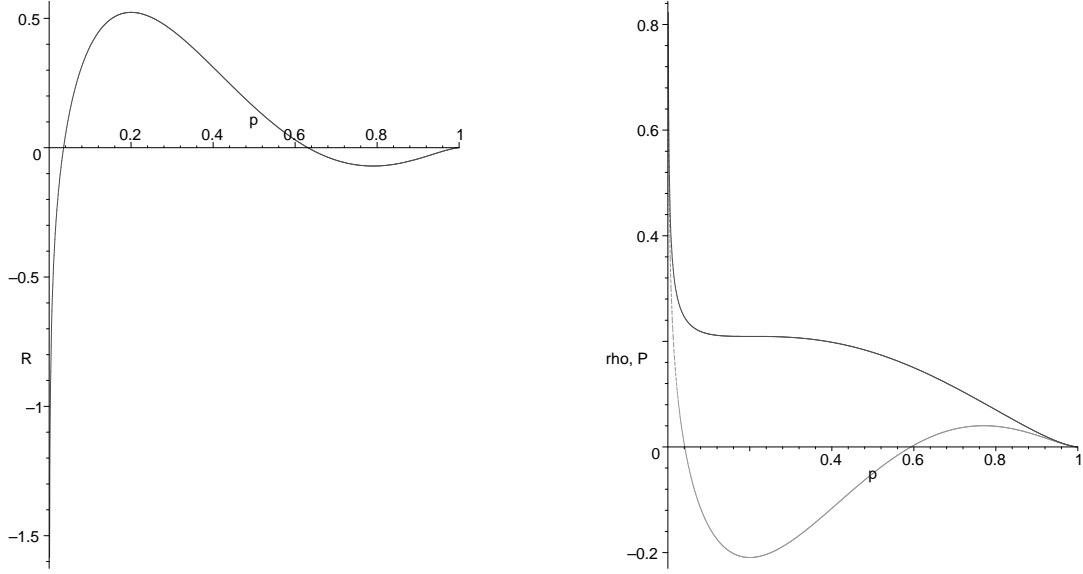

\center{\epsfig{file=cl1so9time01.eps,width=6cm}\hfill 
\epsfig{file=cl1enpres01.eps,width=6cm}\hfill}
\vskip 5mm
\caption{Plot of the Ricci scalar and $\rho(p)$, $P(p)$ (dotted line) 
for the $SO(9)$ 
time-dependent solution.}
\label{enpresscl1}
\end{figure}

The unique timelike solution in the Einstein frame is,
$$
ds^2=-(1-p)^{-7/2} p^{-13/8}{dp^2 \over 4\alpha e^{5\phi_0/2}}
+(1-p)^{-1/6} p^{1/24}\eta_{\mu \nu}
dx^\mu dx^\nu
$$
$$
\phi=\phi_0+\ln (1-p)+{1\over 4} \ln p
$$
with $0<p<1$ where $\eta$ is gauged away since it only 
interchanges $p\rightarrow 1-p$. In the case 
of $SO(9)$ symmetry we have a homogeneous and isotropic 
perfect fluid generated by the dilaton field. 
The energy density 
and pressure give,
$$
\rho(p)={1\over 16}\alpha e^{5\phi_0/2}{(3p+1)^2 (1-p)^{5/2}\over p^{3/8}}
$$
$$
P(\rho)={1\over 16}\alpha e^{5\phi_0/2}(41p^2-26p+1)(1-p)^{3/2} p^{-3/8}
$$
Note then how the fluid is concentrated in the region of the initial 
 cosmological singularity (figure \ref{enpresscl1}).
We see that pressure changes sign accordingly 
to the scalar curvature of spacetime. 
This behavior is reminiscent 
of cosmological models with scalar fields \cite{halliwell} in 4 dimensional 
standard cosmology and in extension to quintessence models which strive
into explaining the present acceleration \cite{quint1}, \cite{quint} 
of our Universe. Note that as 
$p\rightarrow 1$ the dilaton field approaches minus infinity and therefore 
rolls down the potential well with the dilatonic matter dissipating 
in that region. However not surprisingly 
spacetime is not asymptotically flat.   

The last solution of $SO(9)$ symmetry is simply obtained 
by imposing $c=d=0$ (\ref{pope}).
 Strictly speaking it belongs to Class II solutions 
but we include it here for completeness. 
This solution exhibits the asymptotic behavior 
of all Class I solutions of compact proper distance. It reads,
$$
ds^2= p^{-1/8}\left(\eta_{\mu \nu} dx^\mu dx^\nu+{dp^2\over 4\alpha 
e^{5\phi_0/2}p^5}\right)
$$
$$
\phi=\phi_0+{5\over 4}\ln p
$$ 
The above  three static solutions are the unique $SO(9)$ solutions in the 
non-supersymmetric heterotic theory. The important point is that 
these solutions are not asymptotically, flat or  of constant curvature, even 
though the scalar field  rolls down  to a global minimum.

\subsection{Class II solutions}

This case is defined by $\Delta=0$ which implies that
 $d=c(-\gamma\pm  \sqrt{2s})$, $s>0$
and the double root of $f(p)$ is given by $p_1=\pm{c \over \sqrt{2s}}$.
Then we use (\ref{B}) and (\ref{phinu}) 
and performing a change of origin, 
\be
\label{class2}
ds^2=p^{-2\over (D-2)s}e^{2p_1\over (D-2)ps}\left[e^{d\over sp}\left(-dt^2 
+{e^{2p_1\over ps} dp^2 \over {e^{\gamma\phi_0}2\alpha s p^{4+{2\over s}}}}
\right)+
dx_{D-2}^2\right]
\ee
\be
\label{class2dilaton}
\phi=\phi_0 +{\gamma \over s}\ln p +{c-\gamma p_1\over sp} 
\ee
for $p>0$. Note then that the metric (\ref{class2}) is singular and  has one
candidate singularity at the shifted location of the root.

Consider now the 10-dimensional heterotic case where $\gamma=5/2$ i.e. $s=2$. 
We have two static solutions 
characterized by the roots, $p_1=\pm c/2$ and $d=-c/2$, $d=-9c/2$ 
respectively. 
Consider first the couple $p_1=c/2$,  $d=-c/2$. By setting 
$p\rightarrow {|c|p\over 4}$ and taking $d\rightarrow d/c$ 
the solution is given by,
\be
\label{class21}
ds^2=p^{-1/8}e^{{\eta\over 4p}}\left(-e^{-\eta \over p}dt^2 
+e^{\eta\over p} {dp^2\over p^5 e^{5\phi_0/2}4\alpha}+
dx_8^2\right)
\ee
\be
\label{dil21}
\phi=\phi_0+{5\over 4}\ln p-{\eta\over 2p}
\ee
where as before $\eta=c/|c|=\pm1$. 
Not surprisingly for large $p$ (\ref{class21})
asymptotes the $c=d=0$ solution (\ref{pope}).
In the string frame the solution is,
$$
ds^2=p^{1/2}(-e^{-\eta \over p}dt^2 
+e^{\eta\over p} {dp^2\over p^5 e^{5\phi_0/2}4\alpha}+
dx_8^2)
$$

As for Class I solutions (\ref{class21}) 
has a totally different behavior
depending on the sign of $\eta$. Indeed for $\eta=-1$ 
the proper distance in the 
$p$ direction is
finite.
\begin{figure}
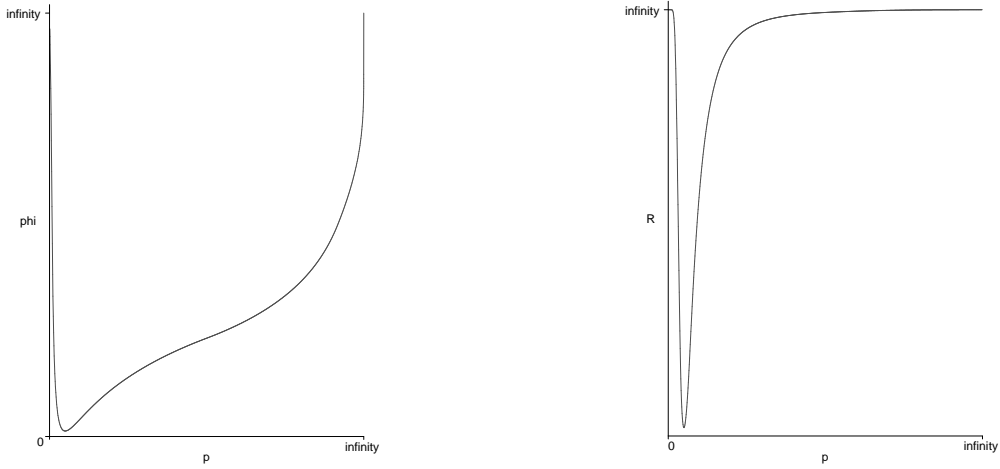

\center{\epsfig{file=cl21phi01.eps,width=5cm}\hfill 
\epsfig{file=cl21ricci01.eps,width=5cm}\hfill}
\vskip 5mm
\caption{Plot of the dilaton
 field $\phi(p)$ and the Ricci scalar curvature $R(p)$ for  the
``compact'' $\eta=-1$ case in Class II}
\label{cl2}
\end{figure}
Hence spacetime is spontaneously compactified in the $p$ direction both 
in the Einstein and in the string frame. Furthermore 
calculation of the Ricci scalar shows that for $\eta=-1$ (\ref{class21}) 
has two naked singularities at $p=0$ and at coordinate infinity,
$$
R={\alpha e^{5\phi_0\over 2}\over 8}
p^{9\over 8}e^{-{5\eta\over 4p}}(45p^2+20\eta p+4)
$$
The dilaton potential (and the Ricci scalar) have a global minimum 
at finite non-zero $p$ where the exponential term takes over the power law 
(see figure \ref{cl2}).

\begin{figure}
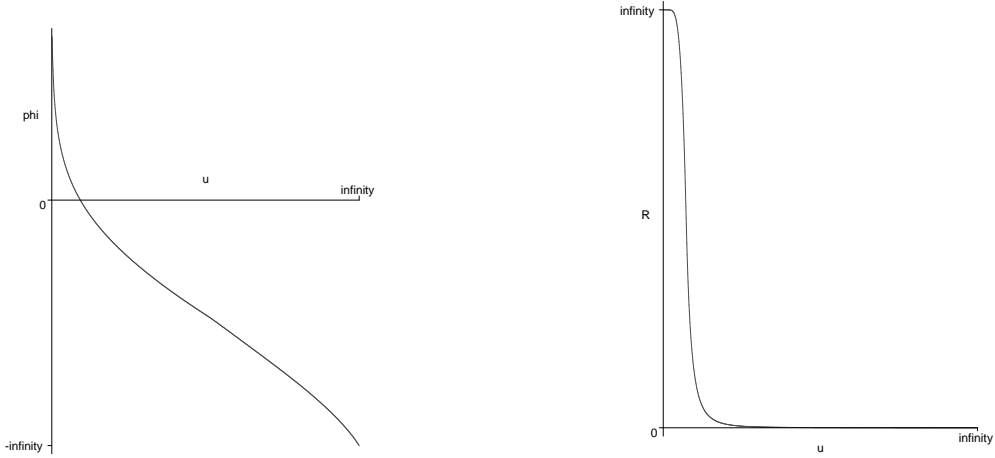

\center{\epsfig{file=cl22phi01.eps,width=5cm}\hfill 
\epsfig{file=cl22ricci01.eps,width=5cm}\hfill}
\vskip 5mm
\caption{Plot of the dilaton field $\phi(u)$ and the Ricci scalar $R(u)$
for $\eta=1$ (infinite proper distance) for Class II.}
\label{cl22}
\end{figure}

However for $\eta=1$ proper distance is infinite in $p$ and accordingly $p=0$
is only a coordinate singularity with spacetime diverging at spatial 
infinity. We coordinate 
transform (\ref{class21}) bringing $p=0$ to infinity, $u=1/p$, and 
(\ref{class21}) gives,
$$
ds^2=u^{1/8}e^{{u\over 4}}\left(-e^{-u}dt^2 
+u e^{u} {du^2\over e^{5\phi_0/2} 4\alpha}+
dx_8^2\right)
$$
$$
\phi=\phi_0+{5\over 4}\ln {1\over u}-{u\over 2}
$$ 
Hence now we have a naked singularity at $u=0$ and at large $u$  dilaton 
matter  tends to  $0$.
Note in  figure \ref{cl22} how the absolute minimum of $\phi$ at minus 
infinity is 
attained for $u\rightarrow\infty$ with the curvature scalar going
accordingly to $0$.

When $p_1=-c/2$ and $d=-9c/2$ we get in a similar fashion,
$$
ds^2=p^{-1/8}e^{{-\eta\over 4p}}\left(-e^{-9\eta \over p}dt^2 
+e^{-11\eta\over p} {dp^2\over p^5 e^{5\phi_0/2}4\alpha}+
dx_8^2\right)
$$
$$
\phi=\phi_0+{5\over 4}\ln p +{9\eta \over 2p}
$$
Again 
for $\eta=1$ the $p$-direction is  compact with the according behavior 
of the curvature tensor whereas $\eta=-1$ yields infinite proper length.  

We note again the similar topological role played by the integration constant 
$c$ and the persisting property, proper distance is compact iff curvature 
diverges at the endpoints of the proper interval.

\subsection{Class III solutions}

In this case we have that $f(p)>0$ for all $p$. We can therefore
anticipate that there will not be any singular points 
for finite values of coordinate $p$ which can therefore 
vary on the entire real line.
This class of solutions is only attained for $s>0$ and the negative 
discriminant is given by $-\Delta={c^2 \psi^2\over s^2}$
where we have set $\psi^2=2s-(d/c+\gamma )^2$ a positive real number. 
We proceed as before using (\ref{B}) and then (\ref{phinu}) to 
find $B$, $\nu$ and $\phi$.
Performing a change of origin, $p-{d+\gamma c\over 2s}\rightarrow p$
to simplify notation we find
the $D$-dimensional Class III metric,
\be
\label{class3}
ds^2={e^{-{2(d+\gamma c)\over s\psi(D-2)}q(p)}\over 
[p^2+{\psi^2\over 4s^2}]^{1\over (D-2)s}} 
\left[e^{-{2d\over \psi}q(p)}
\left(-dt^2+{e^{-{2(d+ \gamma c)\over s \psi}q(p)} dp^2\over 
2\alpha s e^{5\phi_0/2} [p^2+{\psi^2\over 4s^2}]^{4+{1\over s}}}\right)
+dx_{D-2}^2\right]
\ee
$$
\phi=\phi_0+{\gamma\over 2s} \ln \left(p^2+{\psi^2\over 4s^2}\right)+
{\gamma(d+\gamma c)-2cs\over s\psi}q(p)
$$
where $p\in (-\infty, \infty)$ 
and we have set $q(p)=$Arctan$({2ps\over \psi})$. 

We now examine the $D=10$, heterotic $\gamma=5/2$ 
background, whereby we set $p\rightarrow p\psi/4$, we relabel $d/c$ by $d$
and therefore $-9/2<d<-1/2$ for positive $\psi$.  We 
perform the coordinate transformation $\tan q={4p\over \psi}$ 
bringing spatial infinity to a finite value with $q\in (-\pi/2,\pi/2)$.
After some algebra and using (\ref{constants}) we obtain
$$
ds^2=(\cos q)^{1/8}e^{-{d+5/2 \over  8\psi}q}
\left(-e^{-{2d q\over \psi}}dt^2+
{\cos q\, e^{-{3d+5 \over \psi}q}\over 4\alpha e^{5\phi_0/2}} 
dq^2 +dx_8^2\right)
$$
$$
\phi(q)=\phi_0-{5\over 4}\ln( \cos q) +5{d+9/10\over 4\psi}q
$$
Now it is obvious that given the finite interval at which our spacelike 
coordinate $q$ varies that the proper distance in the $q$-direction 
will always be finite.  Hence Class III solutions are always compact in the $q$
direction. Furthermore from the dilaton field we see that spacetime curvature
diverges as $q$ approaches $\pm \pi/2$. 
In the string frame the Class III solution reads,
$$
ds^2=(\cos q)^{-1/2}e^{{2d+1 \over  4\psi}q}\left(-e^{-{2d q\over \psi}}dt^2+
{e^{-{3d+5 \over \psi}q}\over 4\alpha e^{5\phi_0/2}} 
\cos q dq^2 +dx_8^2\right)
$$
and the above characteristics remain true.  
\begin{figure}
\center{\epsfig{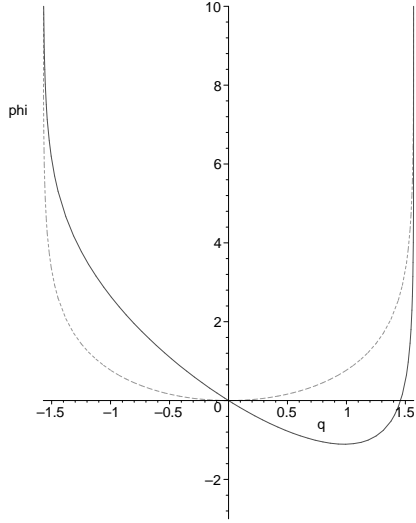}}
\vskip 5mm
\caption{Plot of the dilaton field $\phi(q)$ for $d=-9/10$ (dotted line) 
and $d=-7/2$ for Class III.}
\label{cl3}
\end{figure}
The dilaton potential
 has a unique local minimum for positive $q$ and diverges at the ends 
of the interval. The Ricci scalar is well 
behaved up until we reach $q=\pm \pi /2$ where it explodes exponentially fast. 
Note the particular value $d=-9/10$ which was the value that separated compact from infinite proper distance for the class I solutions. Here it yields an 
even function for the dilaton. For $-9/2<d<-9/10$ we have a local minimum 
which approaches $\pi/2$ as $d\rightarrow -9/2$ (see figure \ref{cl3}) 
and conversely for 
$-9/10<d<-1/2$ the minimum approaches $-\pi/2$. To summarize, although 
Class III solutions do not have candidate singularities at finite values
 of $p$, however, the proper distance in the $p$ direction is finite
 with spacetime exploding at finite proper distance at the endpoints of
 the proper interval. Therefore Class III solutions are also singular.

\section{The critical case and the open string}

When $s=0$ the critical value of $\gamma$ is given (\ref{s}) by 
$\gamma=\sqrt{2(D-1)\over (D-2)}$
and one has to start from (\ref{p}) to obtain 
$$
lnB=(d+\gamma c)^{-1}\int{pdp\over f(p)}
$$
where $f(p)$ is now a first degree polynomial,
\be
\label{f(p)32}
f(p)=p-{c^2\over 2(d+\gamma c)} 
\ee
We label the unique root by, $p_1={c^2\over 2(d+\gamma c)}$
and we anticipate a spacetime singularity at this point. 
The components are  easily integrated from (\ref{p}), (\ref{static22}) and 
(\ref{static23}) yielding,
$$
B(p)=B_0 e^{p\over d+3c/2}|p-p_1|^{c^2\over 2(d+3c/2)^2},\qquad
e^{2\nu}=e^{2\nu_0} B^{{(D-1)\over(D-2)}}|p-p_1|^{d\over d+3c/2},\qquad
e^{\phi}=e^{\phi_0} B^{-\gamma}|p-p_1|^{c\over d+3c/2}
$$
Using (\ref{B}) and (\ref{static1}) we see that the integration 
constants are related by,
\be
\label{constants32}
(d+\gamma c)f(p)=-2\alpha e^{2\nu_0+\gamma\phi_0}|f(p)|
\ee
which states in particular that $d+\gamma c<0$ for static solutions and 
$d+\gamma c>0$ for time-dependent solutions{\footnote {Note that 
$c=d=0$ implies 
from (\ref{constants32}) that $\alpha=0$ and is thus excluded}}. Let us 
start with
the former case. 
Performing a change of origin $p-p_1\rightarrow p$ and considering $p>0$
we obtain the following static solution,
\be
\label{sols}
ds^2=-e^{2p\over (D-2)(d+\gamma c)}p^{2p_1\over (D-2)(d+\gamma c)} \left[
p^{d\over d+\gamma c}\left(-dt^2
+{e^{2p\over d+\gamma c} p^{{2p_1\over d+\gamma c}-2}dp^2 \over 
(d+\gamma c)^2 e^{\gamma\phi_0}}\right)
+dx_{D-2}^2\right]
\ee
\be
\label{dilaton32}
\phi=\phi_0-{\gamma\over d+\gamma c}p+{c-\gamma p_1\over d+\gamma c}\ln p
\ee

Let us now examine the case $D=10$, $\gamma=3/2$ which corresponds to $SO(8)$ 
backgrounds for the Type I string.
As before we can absorb the constant $c$ modulo a sign by setting 
$p\rightarrow p|c|$, putting $\eta=|c|/c$ and relabeling $d\rightarrow d/c$.
The case of $c=0$ was examined in section III. 
For $p>0$ we have from
(\ref{constants32}) that $d+3/2<0$ for $\eta=1$ (or $d+3/2>0$ for $\eta=-1$)
for a spacelike
solution whereas we have $d+3/2<0$ and $\eta=-1$ (or $d+3/2>0$ for $\eta=1$) 
for timelike solutions. 

\begin{figure}
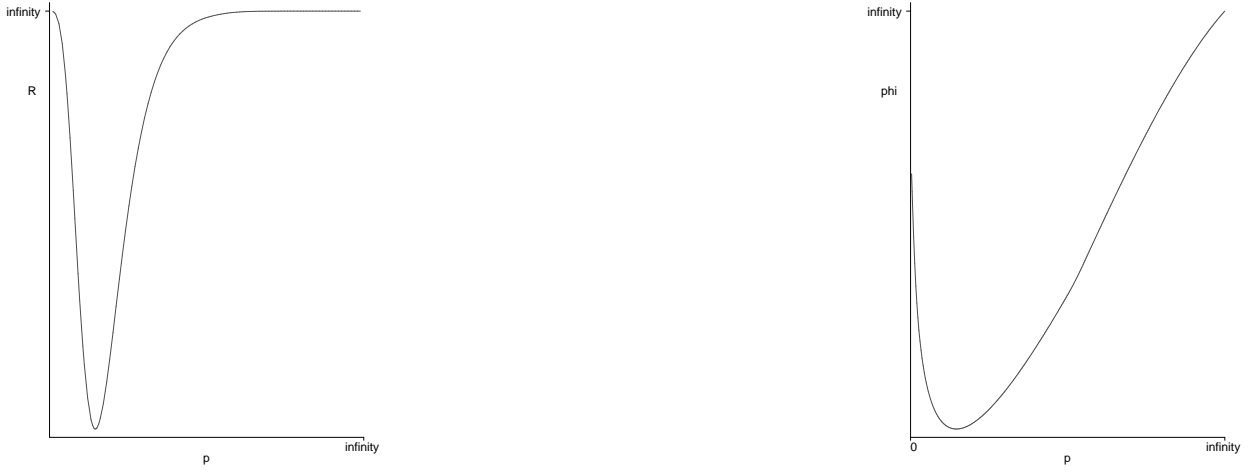

\center{\epsfig{file=clusp101.eps,width=5cm}\hfill
\epsfig{file=clusp201.eps,width=5cm}}
\vskip 5mm
\caption{Plot of the Ricci scalar $R(p)$ and dilaton field $\phi(p)$ 
for $d=-5/2$ and  $\eta=1$ (static case) for $s=0$.}
\label{clusp}
\end{figure}

The $USp(32)$ spacelike solution ($c\neq 0$) written in the Einstein frame is,
\be
\label{sol32}
ds^2=e^{\eta p\over 4(d+3/2 )}p^{1\over 8(d+3/2)^2} \left[
p^{d\over d+3/2 }\left(-dt^2
+{e^{2p\eta \over d+3/2 } p^{{1\over (d+3/2)^2 }-2}dp^2 \over 
(d+3/2)(-\eta) e^{3\phi_0/2} 2\alpha}\right)
+dx_8^2\right]
\ee
$$
\phi=\phi_0-{3\eta\over 2d+3}p+{d+3/4\over (d+3/2)^2}\ln p
$$
where $p>0$ and we have $\eta(d+3/2)<0$.
Quite generically the behavior of solutions (\ref{sol32}) 
near  $p=0$ is controlled by the powers of $p$ whereas the asymptotic
large $p$ behavior is controlled by the exponentials. A simple 
calculation shows that the spacelike solutions are {\it always}
 compact in the $p$
direction.
Then as before we have two naked singularities, one at $p=0$ and one at 
finite proper distance (but at coordinate infinity). So once more we see that 
proper distance is compact iff the curvature diverges at the endpoints 
of the $p$-interval. In the string frame $g_s=e^{\phi/2}g$ 
all the properties discussed go through as
 in the Einstein frame solution (\ref{sol32}).
 To illustrate the above results let us consider 
the explicit example where $d=-5/2$.
In this case the curvature scalar reads,
\be
\label{ex32}
R={\eta\over 16}e^{3\phi_0/2}\alpha e^{9p\eta/4}{36p^2-44\eta p+49\over p^{29/8}}
\ee
Now notice when $\eta=1$ (spacelike case) 
curvature diverges at both ends with the dilaton 
field acquiring a global minimum at the root of the polynomial for
finite $p$ (figure \ref{clusp}).

Proceeding from (\ref{constants32}) the timelike solution reads,
$$
ds^2=e^{\eta p\over 4(d+3/2 )}p^{1\over 8(d+3/2)^2} \left[
p^{d\over d+3/2 }\left(
-{e^{2p\eta \over d+3/2 } p^{{1\over (d+3/2)^2 }-2}dp^2 \over 
\eta (d+3/2) e^{3\phi_0/2} 2\alpha}  +dz^2 \right)
+dx_8^2\right]
$$
where $p$ is a positive timelike coordinate and now $(d+3/2)\eta>0$. 
The change of signature totally changes the asymptotic behavior of the
time-dependent solutions. To illustrate this take again $d=-5/2$, and
note the change in the Ricci scalar (\ref{ex32}) for $\eta=-1$. 
Proper time is always infinite iff $\eta=-1$
and now dilaton matter is concentrated around $p=0$ whereas it  is 
exponentially damped for larger $p$. Accordingly the dilaton scalar rolls down 
the potential well with $\phi\rightarrow -\infty$ as $p\rightarrow +\infty$,
however,  spacetime is not asymptotically flat. Weyl curvature (depending 
on parameter $d$) takes over at large $p$ inducing purely gravitational 
forces. Although the dilaton rolls down the potential 
well, spacetime is not trivial as one could have naively expected.
\begin{figure}
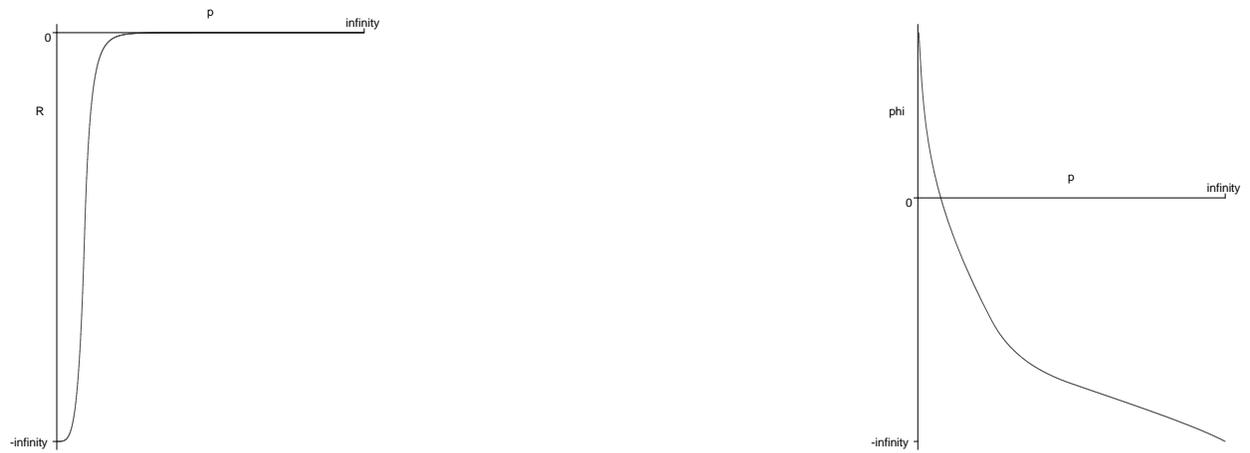

\center{\epsfig{file=clusp301.eps,width=5cm}\hfill
\epsfig{file=clusp401.eps,width=5cm}}
\vskip 5mm
\caption{Plot of the Ricci scalar $R(p)$ and dilaton field $\phi(p)$ 
for $d=-5/2$ 
and $\eta=-1$
(timelike case) for $s=0$.}
\label{cl2usp}
\end{figure}

 It is now instructive to look at 
some asymptotic values of $c$ and $d$. The first case we consider is 
$d=-3/2$, since for this value the polynomial $f(p)=-c^2/2$ is just a 
non-zero constant.
The solution reads,
$$
ds^2=e^{-{p^2\over 4}+3p}(-dt^2+e^{-2p^2}{dp^2\over e^{3\phi_0/2} 4\alpha})
+e^{-p^2/4}dx_8^2
$$
$$
\phi=\phi_0+{3p^2\over 2}-2p
$$
with $p\in R$. Proper distance is again finite and this solution 
explodes at the tips of the proper interval. 

\noindent
{\bf \underline{d=0}}:
This case gives maximal $SO(9)$ symmetry solutions of vanishing Weyl curvature,
$$
ds^2=e^{p\eta\over 6}p^{1 \over 18} (-{dt^2\over -\eta} +dx_8^2)
+{e^{-3\phi_0/2}\over -3\alpha\eta}e^{3p\eta\over 2}p^{-3\over 
2}dp^2
$$
$$
\phi=\phi_0-{\eta p}+{1\over 3}\ln p
$$
For $\eta=-1$ we get the unique spacelike solution of compact proper distance in $p$ and upon making the 
coordinate transformation 
$p={3\over 4} \alpha y^2$ and relabeling the constants 
we recognize the  \cite{dm} solution with $SO(9)$ symmetry.
For $\eta=1$ we obtain the unique timelike solution,
$$
ds^2=-{e^{-3\phi_0/2}\over 3\alpha}e^{3p\over 2}p^{-3\over 
2}dp^2+e^{p\over 6}p^{1 \over 18} dx_9^2
$$
$$
\phi=\phi_0- p+{1\over 3}\ln p
$$
\begin{figure}
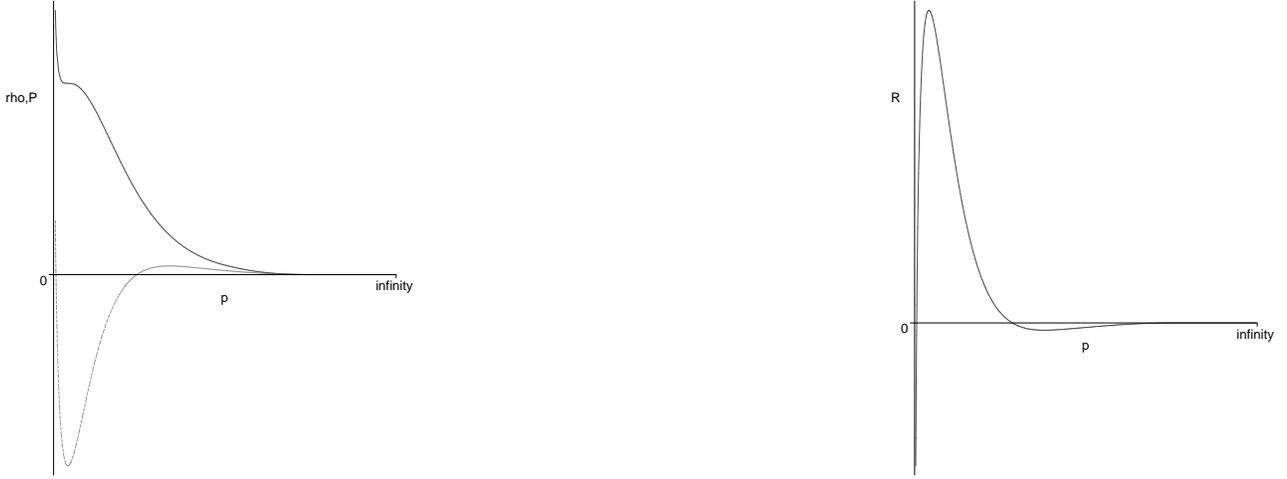

\center{\epsfig{file=clusp601.eps,width=5cm}\hfill
\epsfig{file=clusp501.eps,width=5cm}}
\vskip 5mm
\caption{Plot
of the energy density $\rho(p)$, pressure $P(p)$ (thin line) and 
scalar curvature $R(p)$ for the $d=0$ timelike case ($s=0$). 
The pressure changes sign following the behavior of
spacetime curvature.}
\label{cl3usp}
\end{figure}
This solution (see figure \ref{cl3usp}) 
is reminiscent of scalar field cosmologies studied in \cite{halliwell}. 
As time tends to infinity the dilaton field rolls down the potential
hill obtaining its vacuum expectation value. The energy density and pressure 
of the homogeneous and isotropic dilaton fluid read,
$$
\rho(p)=\alpha e^{3\phi_0/2}
\left({3\over 4}p^2+{1\over 2}p+{1\over 12}\right)e^{-3p/2}p^{-1/2}
$$
$$
P(p)=\alpha e^{3\phi_0/2}
\left({3\over 4}p^2-{1\over 2}p+{1\over 12}\right)e^{-3p/2}p^{-1/2}
$$
and fall exponentially fast to a constant value  for large time $p$. 
Accordingly the Ricci scalar has 
a naked singularity at $p=0$ and for large  $p$ tends to a finite
value. Again we do not have asymptotic flatness however.

To summarize we saw that the static 
type I string solutions of $SO(8)$ or indeed of higher 
symmetry are always of compact proper distance in $p$ and are hence singular
at the endpoints of the interval.
The timelike solutions exhibit a rather interesting behavior, namely matter is 
heavily concentrated around the origin and is exponentially dumped as we move
away with the dilaton field rolling down the potential hill.

\section{Treating the general problem}

Let us now go back to the general problem of solving (\ref{einstein2D}) 
under the light of 
the one-dimensional solutions. The question we are
tackling in this section 
is the existence of 2-dimensional solutions of (\ref{einstein2D}). Put in 
a different way we are asking if the general one-dimensional solutions we 
obtained are the unique solutions of (\ref{einstein2D}){\footnote{Discussions 
with David Langlois and Maria Rodriguez-Martinez have considerably improved
our understanding of this section}}. If this were
the case we would obtain an extension of Birkhoff's theorem,
in the sense that there would exist a local timelike or spacelike extra Killing
vector for our two dimensional system making thus solutions 1-dimensional.

Let us start by analyzing the field equations.
We note  using (\ref{wave11D}) in (\ref{wave13D}) that we
obtain the {\it linear-like equation},
\be
\label{phi}
B\phi_{uv}+{1\over 2}(B_u\phi_v+B_v\phi_u)+\gamma B_{uv}=0
\ee
We hence set without loss of generality,
\be
\label{usa}
\phi=\tilde{\phi}-\gamma lnB
\ee
and (\ref{phi}) simplifies to,
\be
\label{phitilde}
(B\tilde{\phi_u})_v=-(B\tilde{\phi_v})_u
\ee
where $\tilde{\phi}$ is now the unknown field. 
Furthermore let us define,
\be
\label{chi}
2\chi=2\nu+\gamma\tilde{\phi}-{\gamma^2\over 2}lnB
\ee
upon which the system (\ref{einstein2D}) simplifies to,
\bml
\label{einste2}
\bea
B_{uv} &=& 2\alpha B^{-(s+1)} e^{2\chi}\label{wa11}\\
(2\chi)_{uv} &=& 
-2\alpha(s+1) B^{-(s+2)} e^{2\chi}
-{1\over 2}\tilde{\phi}_{u}\tilde{\phi}_{v} 
\label{wa12}\\
\tilde{\phi}_{uv} &=& -
{1\over 2B}(\tilde{\phi}_u B_v+\tilde{\phi}_vB_u\label{wa13})\\
(2\chi)_{u}-\left[ln(B_{u})\right]_{u}&=& {B \over 2B_u}\tilde{\phi}_u^2
\label{i11}\\
(2\chi)_{v}-\left[ln(B_{v})\right]_{v} &=& {B \over 2B_v}\tilde{\phi}_v^2
\label{i12}
\eea
\eml
The independent variables are now $B$, $\tilde{\phi}$ and $\chi$.
Note that $\alpha$, $\gamma$ and $D$ appear only in the wave equations 
(\ref{wa11}) and (\ref{wa12}) where we remind the reader that
$s={\gamma^2\over 2}-{D-1\over D-2}$. 
Furthermore since (\ref{wa12}) results from the 
other field equations, (\ref{wa11}) is the only equation depending
on the parameters of the problem. Note also that as long as 
 $\alpha\neq 0$ we have  $B_{uv}\neq 0$
as can be seen from (\ref{wa11}) and therefore the $B$ field cannot be 
trivial. By the redefinition of the field 
components we have  
effectively absorbed the dilaton potential in our new variables. 
The system (\ref{einste2}) corresponds to an energy-momentum tensor 
consisting of a scalar field with cosmological
constant $\alpha$.\footnote{For $\gamma=1/6$ and $D=10$ we get from 
${1/(D-2)-\gamma^2/2}=1/(n-2)$ an $n=11$-dimensional spacetime as one 
would expect from KK type compactification}

Now from (\ref{wa11}) we can read off,
\be
\label{chieq}
e^{2\chi}={1\over 2\alpha} B_{uv} B^{s+1}
\ee
and then replace it in the integrability conditions (\ref{i11}) and 
(\ref{i12}),
\be
\label{o1}
\tilde{\phi}_u^2=2{B_u B_{uvu}\over B B_{uv}}+2(s+1)
{B_u^2\over B^2}-2{B_{uu}\over B}
\ee
\be
\label{o2}
\tilde{\phi}_v^2=2{B_v B_{vvu}\over B B_{uv}}+2(s+1)
{B_v^2\over B^2}-2{B_{vv}\over B}
\ee

Now in principle we have two unknown functions $B$ and $\tilde{\phi}$ 
and four equations to solve. 
Which of these equations are independent? Suppose we make use only of 
(\ref{o1}) and (\ref{o2}); can we obtain (\ref{wa12}) and (\ref{wa13})?
Differentiating (\ref{o1}) and (\ref{o2}) with respect to $v$ and $u$ 
respectively we obtain,
$$
2B\tilde{\phi}_{uv}+B_v\tilde{\phi}_u-2{B_u\over\tilde{\phi}_u} {\cal {A}}=0
$$
$$
2B\tilde{\phi}_{uv}+B_u\tilde{\phi}_v-2{B_v\over\tilde{\phi}_v} {\cal {A}}=0
$$
which are symmetric under $u\leftrightarrow v$. The functional,
$$
{\cal {A}}=(lnB_{uv})_{uv}+(s+1)\left(2{{B_{uv}}\over B} 
-{B_uB_v\over B^2}\right)
$$
is symmetric in $u$ and $v$.
Combining these two equations 
we get,
$$
\left(1-{B_v\tilde{\phi}_u \over B_u \tilde{\phi}_v} \right)
(2B\tilde{\phi}_{uv}+B_u\tilde{\phi}_v+B_v\tilde{\phi}_u)=0
$$
and 
$$
\left(1-{B_v\tilde{\phi}_u \over B_u \tilde{\phi}_v} \right)\left((2\chi)_{uv} 
-2\alpha(s+1) B^{-(s+2)} e^{2\chi}-
{1\over 2}\tilde{\phi}_{u}\tilde{\phi}_{v}\right) =0
$$
We recognize as factors (\ref{wa13}) and (\ref{wa12}) 
which therefore result from (\ref{o1}) and (\ref{o2}) as long 
as we {\it don't} have,
\be
\label{isa}
{B_v\tilde{\phi}_u = B_u \tilde{\phi}_v}
\ee
Now to what extend is (\ref{isa}) a relevant equation?
Using (\ref{wa13}) and (\ref{isa}) results to $B=B(U(u)+V(v))$ which
leads to a one dimensional solution studied in the 
previous section. 
Indeed we remind the reader that any function of $U+V$ can 
be reduced to a $z$-dependent function using coordinate transformations 
(\ref{kruskal}). This is due to the fact that we have imposed $SO(D-2)$ 
symmetry hence the two remaining dimensions admit $2$-dimensional 
 conformal invariance (\ref{confD}). 
Therefore any strictly 
two-dimensional
solutions will not obey (\ref{isa}) and hence we deduce that 
(\ref{o1}), (\ref{o2}) 
yield the wave equations (\ref{wa12}) and (\ref{wa13}). Conversely the 
argument follows through in exactly the same way. Hence we deduce 
the following: Equations (\ref{o1}), (\ref{o2}) are equivalent to 
equations (\ref{wa12}) and (\ref{wa13}) for {\it two-dimensional solutions}.

Hence given $B$ we can evaluate $\chi$ from (\ref{chieq}).
Then from (\ref{o1}), (\ref{o2}) 
evaluate $\tilde{\phi}_u$ and $\tilde{\phi}_v$.
Two conditions have then to be met in order to obtain
 a two-dimensional solution.
First of all we must not have (\ref{isa}) for then the solution can be 
coordinate transformed to a 1-dimensional solution. Secondly 
$\tilde{\phi}_u$ and $\tilde{\phi}_v$ have to be differentiated with respect 
to $v$ and $u$ respectively 
and must yield the same result (equivelantly $\tilde{\phi}$ is a 0-form).

So finding two-dimensional solutions seems a difficult task.
However there is a way which gets around this difficulty. 
Indeed note that integrability conditions (\ref{o1}) and (\ref{o2}) 
involve the {\it square} of the derivatives of $\tilde{\phi}$ in $u$ and $v$. 
Hence a way to circumvent 
(\ref{isa}) is to take $\phi_u$ and $\phi_v$ effectively constant{\footnote{i.e. $\tilde{\phi}\sim U-V$}}  {\it
but} of opposite sign.

Indeed let us apply the above algorithm with $B=e^{A(U(u)+V(v))}$
which again represents a general functional of $U+V$. From
(\ref{chieq}) we obtain the $\chi$ field,
$$
e^{2\chi}={1\over 2\alpha}(A''+A'^2)U'V'B^{s+2}
$$
and from (\ref{o1}),
\be
\label{peter}
\tilde{\phi_u}^2={2U'^2 \over A''+A'^2}[A'A'''+(2+s)A''A'^2+(s+1)A'^4-A''^2]
\ee
((\ref{o2}) yields the analogous equation for ``$v$'') 
and {\it prime} stands for
the derivative with respect to $U+V$.
A simple example to consider is $A=U+V$, \cite{davidmaria}. 
Then it is straightforward
to get,
$$
e^{2\chi}={1\over 2\alpha}U'V'e^{(s+2)(U+V)}
$$  
and
$$
\tilde{\phi}=\pm\sqrt{2(s+1)}(U-V)
$$
Note then the $U-V$ dependence of $\tilde{\phi}$ and consequently 
that (\ref{isa}) is not true and that the solutions are 
two dimensional. Had we taken the same sign for 
the derivatives of $\tilde{\phi}$ we would have got 
a static solution. This sign ambiguity in (\ref{o1}) and (\ref{o2}) 
breaks the unicity argument of Birkhoff's theorem. Obviously if 
$\tilde{\phi}$
were a constant this sign ambiguity would not have been possible.
\footnote{The breakdown of Birkhoff's theorem in this context was 
mentioned in \cite{bir} where the authors refer to a 
forthcoming publication (see references within) discussing 
this.}

In the generic case of arbitrary $A(U+V)$ other 2-dimensional solutions can be 
obtained in the following way. From (\ref{peter}) it is clear that we
must not have $\phi_u$ be a function of $A(U+V)$. Again using (\ref{kruskal}) 
this would just imply staticity. Therefore $A$ is 
constrained by the following differential equation,
\be
\label{ruth}
k^2=2{A'A'''+(2+s)A''A'^2+(s+1)A'^4-A''^2 \over A''+A'^2}
\ee
where $k$ is an arbitrary constant. Obviously 
for such $A'$ we get implicitly a 2-dimensional solution with 
$\tilde{\phi}=\pm k(U-V)$. 
 Let us analyze here for simplicity 
the example $A=U+V$ \cite{davidmaria}. 
With the coordinate transformation (\ref{kruskal})
the solution simplifies to,
$$
ds^2={1\over 2\alpha}e^{{\gamma^2}z}
e^{\gamma\eta \sqrt{2(s+1)}t}(-dt^2+dz^2)+e^{2z\over D-2}dx_{D-2}^2 
$$
$$
\phi=\phi_0-\eta\sqrt{2(s+1)}t-\gamma z
$$
The first thing to note is that this solution is valid as long as 
$\gamma^2> {2\over D-2}$. Hence in $D=10$ the critical point is $\gamma=1/2$
where the solution is again static and belongs to Class I. 
For our cases of interest the solution
is then well defined. Furthermore and most importantly this solution is 
everywhere regular. Just like 
the case of thick domain walls \cite{filipe} or global vortices \cite{ruthgl} 
we see that the solution is regular
once we relax the staticity requirement. Indeed the above solution 
has the same general 
form as a thick planar domain wall solution \cite{filipe} by 
taking the coordinate transformation,
$$
U={1\over 2(\gamma-\eta\sqrt{2(s+1)})}(z-t),\qquad V=
{1\over 2 (\gamma+\eta\sqrt{2(s+1)}) }(z+t)
$$
where we have kept the labels for our coordinates as $z$ and $t$.
The solution reads,
\be
\label{dwall}
ds^2=e^{\gamma z}(-dt^2+dz^2)+e^{2\eta t\sqrt{2(s+1)}\over D-2}
e^{\gamma z}dx^2_{D-2}
\ee
$$
\phi=\phi_0-z
$$
We see that the scalar field is now
static but spacetime is not. Note however the absence of the event 
horizon which caracterises the fact that the scalar field \cite{filipe} 
is a topological
soliton.

 There is the case of $\alpha=0$ we can study completely 
following the work of Tabensky and Taub \cite{tabensky}. The
integrability of the free scalar field case yields useful insight for
the unicity of the solutions we have been discussing here.

\subsection{The SUSY limit $\alpha=0$}

The case of $\alpha=0$ amounts to switching off the 
Liouville potential in our action (\ref{actionD}). 
This case was studied in 4 dimensions by Tabensky and Taub
\cite{tabensky} for a stiff perfect fluid source (where pressure is equal
to energy density). We sketch the extension to $D$ dimensions here which
is trivial once we have taken the metric in the form
(\ref{einstein2D}). 
Explicit 
examples can be found in the original paper as well as more recently in 
\cite{blum} for a T-dual version of the Sugimoto model \cite{sugimoto}.

The B equation 
(\ref{wa11}) with $\alpha=0$ 
is just the two-dimensional wave equation which has general
solution,
$$
B=F(u)+G(v)
$$
with $F$ and $G$ arbitrary functions.
 Substituting the solution of $B$ into
(\ref{einste2}) yields,
 
\bml
\label{eins2}
\bea
(2\chi)_{uv} &=& 
-{1\over 2}\tilde{\phi}_{u}\tilde{\phi}_{v} 
\label{w121}\\
\tilde{\phi}_{uv} &=& -
{1\over 2(F+G)}(\tilde{\phi}_u G'+\tilde{\phi}_v F'\label{w31})\\
(2\chi)_{u}-{F''\over F'}&=& {F+G \over 2F'}\tilde{\phi}_u^2
\label{i1111}\\
(2\chi)_{v}-{G''\over G'}&=& {F+G \over 2G'}\tilde{\phi}_v^2
\label{i1121}
\eea
\eml 
Upon making the coordinate transformation, 
$$
(u,v)\rightarrow(F(u),G(v))
$$
the wave equation for $\tilde{\phi}$ reduces to,
\be
\tilde{\phi}_{FG} = -
{1\over 2(F+G)}(\tilde{\phi}_F+\tilde{\phi}_G \label{a2})
\ee 
Now (\ref{a2}) is recognised as an Euler-Poisson-Darboux equation (see
 for instance \cite{ames})
\be
\label{epd}
u_{xy}+{N\over x+y}(u_x+u_y)=0
\ee
which has general solution,
$$
u(x,y)={\partial^{N-1}\over \partial(x+y)^{N-1}}\left({\Phi(x)+\Psi(y)
\over (x+y)^N }\right)
$$
when $N$ is a positive integer. In our case of interest, where $N=1/2$, 
the equation  can be interpreted as a 2 dimensional wave equation 
in cylindrical coordinates. This can be achieved
 by restoring time and space-like coordinates{\footnote {In the 
original analysis \cite{tabensky}, the authors used the Riemmann-Voltera 
method involving hypergeometric functions}},
$$
F=r-t,\qquad G=t+r
$$
to get,
\be
\label{cylindricalwave}
\tilde{\phi}_{rr}+
{1\over r}\tilde{\phi}_r-\tilde{\phi}_t=0
\ee
The general solution is given by means of a variety of integral 
representations (see Copson \cite{copson} for 
a beautiful derivation using complex analysis) here we choose 
Poisson's original formula,
$$
\tilde{\phi}(t,r)=\int_0^\pi \Phi(t+r\cos\psi)d\psi+
\int_0^\pi \Psi(t+r\cos\psi)\log (r \sin^2\psi) d\psi 
$$
with $\Phi$ and $\Psi$ arbitrary $C^2$ functions. 

Hence given $\Phi$ and $\Psi$ we can find $\tilde{\phi}$ and 
then integrate once (\ref{i1111}) and (\ref{i1121}) in order to find 
the $\chi$ field. 

Let us now consider the question of unicity of the field equations
(\ref{einstein2D}). 
The following argument based on the above analysis  leads us to postulate that 
the one dimensional solutions (section III and IV) along 
with the 2 dimensional solutions (\ref{ruth}) constitute 
the general solution to the field equations (\ref{einstein2D}).  
From  the general solution  for 
the case of $\alpha=0$ above 
we saw that  component $B$ verifies a two dimensional 
wave equation hence $B=U+V$. Note then 
that the Taub planar solutions in the vacuum \cite{taub} i.e. in the absence
of a scalar field, also admit as general solution $B=U+V$ {\it in the same 
coordinate system}. On the other hand 
in the case of a bulk of constant curvature, a cosmological constant 
\cite{prc}, the $B$ component verifies a non-homogeneous wave equation 
and it turns out that $B=B(U+V)$. We remind the 
reader that under (\ref{kruskal}) this simply means that $B$ is a static 
field. 
In the case under consideration we showed that the field equations 
(\ref{einstein2D}) could be transformed into (\ref{einste2}), the 
case of a cosmological constant with a scalar field. Therefore by symmetry
one is tempted to postulate that the general solution for $B$ is 
likewise $B=B(U+V)$ just like in the presence of a cosmological constant. 
Under this assumption and noting coordinate transformations 
(\ref{kruskal}) we saw that either we would 
get the one dimensional solutions described in sections III, IV and V
or the 2-dimensional solutions implicitely given by (\ref{ruth}) in this
section.

\section{Conclusions}

Starting from a $D$ dimensional 
spacetime admitting $D-2$ planar symmetry we  
derived in section IV, V the general static or time-dependant solutions. 
Furthermore in section VI we analysed the field equations 
and found implicitely 
a class of two dimensional solutions given by (\ref{ruth}). 
An example of these has recently been 
given in \cite{davidmaria}. Having generalised to $D$ dimensions 
the general solution for $\alpha=0$ (no Liouville potential) \cite{tabensky} 
we conjectured in the end of section VI 
that the solutions we found constitute the unique solutions
to (\ref{einstein2D}). 

For the 1-dimensional case i.e. where the fields are locally 
static or time dependant we found three 
classes of solutions classified by the type of roots of a second degree 
polynomial ($s\neq 0$). Hence Class I solutions involved two distinct roots
for $f(p)$, Class II one double root etc. The locations of the roots of $f(p)$
stood for candidate singularities of spacetime. 
We applied these solutions as gravitating 
backgrounds to  $D=10$ non-supersymmetric string theories, in particular 
the open Type I theory with gauge group $USp(32)$  and the 
heterotic theory with cosmological constant and gauge group 
$SO(16)\times SO(16)$. On passing we obtained the general $SO(9)$ maximal
symmetry solutions recently discussed in \cite{dm}. We saw how 
the type I theory backgrounds, $s=0$, involved a critical value
for  the gravitational field which restricted considerably the possible 
solutions since then $f(p)$ was linear. 
For example whereas the heterotic string admits 3 static and 
1 time dependent solutions of maximal $SO(9)$ symmetry 
the open string admits solely 1 static and 
1 time dependent solution.

All one dimensional solutions depended on two integration 
constants $c=0,-1,1$ and $d$. The former originated from the scalar field and 
was of a topological nature whereas the latter was related to  Weyl 
curvature and $d=0$ simply meant that we had a conformally 
flat spacetime. In the table above we have gathered our 
results for the static heterotic string solutions.

 \vspace*{5mm}
\noindent
\renewcommand{\arraystretch}{1.1}
\begin{tabular}{||c||c |c||c|c|c|c||}\hline \hline
\multicolumn{7}{||c||}{\bf 
Static backgrounds for the non-susy heterotic string, 
$\gamma=5/2$, $D=10$}\\ 
\multicolumn{7}{||c||}{}\\ 
\hline
 & $\bf d$ & $0$ & $d\leq -9/2$ &  $-9/2< d< -1/2$ & $-1/2 \leq d<0$ &$d>0$\\
 & &Class I&Class I, II& &Class I, II&Class I\\
\hline \hline
$\bf c$ & & $\bf SO(9)$ & \multicolumn{4}{c||}{$\bf SO(8)$}\\
 & &  & \multicolumn{4}{c||}{}\\
\hline
 & \itshape Sect.  & Asymptotic  & 
\multicolumn{3}{c|}{Black}& Naked \\
$0$ & \itshape III & Solution & \multicolumn{3}{c|}{Hole}& Singularity\\
 & & &  \multicolumn{3}{c|}{$\infty\leftarrow R \rightarrow 0$}&
{$\infty\leftarrow R \rightarrow 0$}\\
\hline 
 & \itshape Sect. & infinite & compact & compact  & 
\multicolumn{2}{c||}{infinite} \\
$1$ &  \itshape IV & proper dist. & proper dist. &  proper dist. &
\multicolumn{2}{c||}{proper distance} \\
 & & {$\infty\leftarrow R \rightarrow 0$}& 
{$\infty\leftarrow R \rightarrow \infty$}& Class & 
\multicolumn{2}{c||}{$\infty\leftarrow R \rightarrow 0$}\\
\cline{1-4}\cline{6-7}
& \itshape Sect. & compact  & infinite  & III & 
\multicolumn{2}{c||}{Compact } \\
$-1$ & \itshape IV & proper dist.  & proper dist. & & 
\multicolumn{2}{c||}{proper dist.} \\
 & & $\infty\leftarrow R \rightarrow \infty$ & 
$\infty\leftarrow R \rightarrow 0$ & 
{$\infty\leftarrow R \rightarrow \infty$} & 
\multicolumn{2}{c||}{{$\infty\leftarrow R \rightarrow \infty$}} \\
\hline \hline
\end{tabular}
\vspace*{5mm}

The general characteristics of the 1-dimensional 
solutions are as follows: Firstly they are all singular. When $c=0$ the 
singularity could sometimes be censored by a horizon yielding in the 
static case black hole solutions first studied by \cite{cai}. For $c=\pm1$
in certain 
cases we found that  proper distance in the independent spacelike 
variable was actually finite in agreement with the $SO(9)$ solutions of 
\cite{dm}. Then the topology of spacetime is an interval times a nine 
dimensional manifold. Furthermore we found that compactness 
was equivelantly 
related to  singular behavior of the curvature tensor. Indeed
spacetime is always singular at the endpoints of the compact proper interval. 
On the contrary when proper distance is infinite then we had a naked 
singularity at the origin with dilaton matter smoothly going to zero
at large distances or late times. Hence when proper distance is infinite 
the dilaton field will always roll down the potential well towards 
minus infinity where the Liouville 
potential acquires its global minimum. We can actually assert that if 
the dilaton does not roll to its vacuum then proper distance has to be
 compact.
Interestingly the compact $SO(9)$ solutions discussed in detail in \cite{dm}
have effectively not only a 9-dimensional behavior but also 
the dilaton rolls to 
minus infinity at one of the endpoints of the interval. Hence one 
anticipates that this solution
should be classically stable against one of the remaining $SO(9)$ solutions.
Another argument in favor of its stability is that the same type of solution 
appears for the Type I string and there it is the unique static $SO(9)$
solution \cite{dm}.

We should also stress the fact that 
none of the solutions we found are asymptotically flat or of constant 
curvature. 
This seems surprising. Indeed as we noted above 
the scalar field typically will roll to the global minimum of the potential. 
However following the works 
of Poletti and Wiltshire for dilatonic black holes \cite{wiltshire} the
non-asymptotic flatness  
is related to the fact that the potential under question does not acquire its 
minimum at a finite value for $\phi$. Hence even if we allow the dilaton 
to reach its vacuum value, Minkowsky or constant curvature spacetime is 
not a gravitational background  for non-supersymmetric string theories 
at least at the classical level we are considering. 

Let us now turn to the singular nature of the 1-dimensional solutions. It is 
known from studies of gravitating topological defects 
\cite{vil}, \cite{sikivie} that the singular nature of a thick domain wall
\cite{filipe} or of a global vortex \cite{ruthgl} is due to the 
fact that we impose a static spacetime. Indeed on allowing spacetime 
to be space and time dependent global vortex and thick domain wall 
spacetimes are everywhere regular. Exactly the same thing occurs here. 
We saw that a two dimensional solution \cite{davidmaria} is everywhere 
regular and can be transformed  in such a way to acquire 
the form of a thick domain wall solution (\ref{dwall}). By
this we mean that the scalar field is static with 
the spacetime metric having an exponential time dependence and a conformal
space dependent factor. However here 
we have no horizon since the potential is not of a domain wall type,
acquiring a non degenerate discrete set of minima.

The solutions we have obtained 
can be used as background solutions incorporating the motion of brane 
Universe type wall much in a generalized context of \cite{chamblin} 
(for a recent discussion with a two dimensional background 
see \cite{davidmaria}). 
Indeed these solutions are relevant to cosmological perturbations,
 brane cosmology in a non-constant curvature background 
and also to the radion related issues \cite{cgr}. This setup 
is a generalization of a constant curvature spacetime since a Liouville 
potential closely resembles a cosmological constant. In this case 
we can indeed picture the bulk as a fluid of dark matter flowing through
the brane Universe. Thus comes about 
the question of Birkhoff's theorem and its relevance to brane cosmology. 
It was proven recently that Birkhoff's theorem applies in the case of a 
spacetime of constant curvature admitting a $D-2$ spherical planar or 
hyperboloidal symmetry \cite{prc}. 
The unicity then implies Kottler's solution (topological black hole) 
\cite{kottler} as the 
unique brane cosmology background. To understand the essence of this let us 
step back to usual 4-dimensional cosmology. On solving 
the FLRW equations one finds one physical degree of freedom, the expansion
rate of the Universe which is related to energy density and pressure.
When one considers brane cosmology in 5 dimensions
our Universe is a timelike hypersurface evolving in a 5-dimensional spacetime.
With the addition of one dimension one would naively expect that we would 
obtain another physical degree of freedom from the field equations.
 However Birkhoff's theorem does not allow this.
Indeed two dimensional conformal symmetries 
ensure that we have an extra Killing vector which then implies 
only one physical degree of freedom, the wall trajectory, alias the expansion 
rate as witnessed by a four dimensional observer. However as we saw scalar 
field matter breaks this unicity theorem and two dimensional solutions 
exist. This fact is independent of the presence of the potential; indeed 
it would seem that a potential rather 
restricts the possible solutions rather than enhances them.
Could there exist potentials so that an extension to Birkhoff's theorem 
would hold? This seems unlikely since a Liouville type potential is 
a natural generalization of a cosmological constant. 

Let us end on  higher order corrections from the coupling constant expansion. 
For instance in the case of the 
Sugimoto model \cite{sugimoto} and also for the Sagnotti \cite{sagnotti} 
non supersymmetric string model
the potential to consider at one loop and in the  
Einstein frame is $V(\phi)=\alpha_1 e^{3\phi/2}
+\alpha_2 e^{5\phi/2}$ with $\alpha_1$ and $\alpha_2$ positive constants. 
Note now the 
interplay between the critical value $\gamma=3/2$ and $\gamma=5/2$.
Firstly it is simple to show that for  $\alpha_1$ and $\alpha_2$ positive 
no maximally symmetric solution exists. Secondly on a brief 
analysis of the field equations  it is not evident 
that even $SO(9)$ solutions exist with such a potential. We hope to be 
reporting progress on this and further issues in the near future.

\section{Acknowledgments}
During the course of this work it has been very instructive to discuss with
my colleagues in the IFT. I would like to thank David Fairlie for his 
encouragement and motivation during his visit in Florida. Special thanks go 
to Peter Bowcock and Ruth Gregory for continuous support and fruitful 
discussions. It is also 
a great pleasure to thank Emilian Dudas and Jihad Mourad for discussions. 
Furthermore
 it was a great pleasure to discuss with Michael Joyce, David Langlois, 
Maria Rodriguez-Martinez during my visit in France and Cedric Deffayet
during the PASCOS meeting. I finally thank Emilian Dudas, Ruth Gregory and 
Jihad Mourad for comments on the manuscript. CC is supported by
the DOE grant number DE-FG02-97ER41209.

\end{document}